\documentclass[aps,prl,reprint,superscriptaddress,nofootinbib]{revtex4-2}

\usepackage[T1]{fontenc}
\usepackage[utf8]{inputenc}
\usepackage{amsmath,amssymb,amsthm,mathtools}
\usepackage{braket}
\usepackage{bm}
\usepackage{hyperref}
\usepackage{cleveref}
\usepackage{graphicx}
\usepackage{xcolor}
\usepackage{enumitem}
\usepackage{appendix}
\usepackage{tikz}
\usetikzlibrary{arrows.meta,positioning,fit,calc}

\graphicspath{{./images/},{./imagesAppendix/}}

\hypersetup{colorlinks=true,linkcolor=blue,citecolor=blue,urlcolor=blue}

\newcommand{\pL}{p_{\rm L}}
\newcommand{\pth}{p_{\rm th}}
\newcommand{\dt}{\Delta t}

\newcommand{\laml}{\lambda_{\rm low}}
\newcommand{\lamh}{\lambda_{\rm high}}

\newcommand{\SM}{SM}

\newcommand{\prlsection}[1]{{\em #1.---}}

\definecolor{KB}{rgb}{0.6,0.1,0.8}
 
\definecolor{THc}{rgb}{0.9,0.3,0.2}

\newcommand{\idg}[1]{{\bfseries #1)}}

\newcommand{\subfigimg}[3][,]{%
	\setbox1=\hbox{\includegraphics[#1]{#3}}%
	\leavevmode\rlap{\usebox1}%
	\rlap{\hspace*{2pt}\raisebox{\dimexpr\ht1-0.5\baselineskip}{{\bfseries \large\textsf{#2}}}}%
	\phantom{\usebox1}%
}

\begin{document}

\title{%
Exponential logical-error reduction in quantum memories via optimal syndrome-measurement timing
}

\author{Tobias Haug}
\affiliation{Quantum Research Center, Technology Innovation Institute, Abu Dhabi, United Arab Emirates}

\author{Kishor Bharti}
\affiliation{Joint Center for Quantum Information and Computer Science, NIST/University of Maryland, College Park, MD 20742, USA}
\affiliation{Department of Computer Science and Institute for Advanced Computer Studies, University of Maryland, College Park, Maryland 20742, USA}

\author{Leandro Aolita}
\affiliation{Quantum Research Center, Technology Innovation Institute, Abu Dhabi, United Arab Emirates}

\date{\today}

\begin{abstract}
Syndrome-measurement timing is usually treated as a fixed clock cycle of a quantum error-correcting code. For quantum memories, however, the inter-round interval is itself an optimizable control parameter: measuring too rarely allows idling errors to accumulate, whereas measuring too often introduces
measurement-induced faults. We propose a phenomenological logical-noise model for this trade-off and analytically show that the optimal syndrome-measurement interval is inversely proportional to the code distance and that this produces an exponential reduction of logical-error rates in the distance relative to constant-interval schedules. 
Furthermore, for time-dependent idling noise, we develop an adaptive timing strategy based on the measured syndrome activity that outperforms every fixed-interval protocol, with largest gains for short but strong noise bursts. Simulations of rotated surface-code memories with matching decoding validate the phenomenological model, the distance-dependent optimum, and the adaptive-strategy improvement. 
Moreover, with the experimental noise parameters reported by Google in Nature {\bfseries 638} (2025), our model predicts reductions in logical-error rates per unit time of up to \(40\%\).

\end{abstract}

\maketitle

 \let\oldaddcontentsline\addcontentsline%
\renewcommand{\addcontentsline}[3]{}%

\prlsection{Introduction}
Fault-tolerant quantum computers rely on repeated syndrome measurements to convert physical errors into syndrome bits %
that a decoder can use to infer a recovery operation~\cite{lidar2013quantum,terhal2015quantum,spencer2026quantum}. 
The rate at which these measurements are performed is a key operational parameter.
For logical fault-tolerant processors, syndromes are usually measured in quick succession to set a fast %
clock-rate of logical operations.
In contrast, for fault-tolerant quantum memories, whose purpose is to preserve quantum states for as long as possible, syndrome intervals must be more carefully balanced~\cite{pattison2021improved,marton2025optimal,maurya2025synchronization,mundada2026heterogeneous}: 
If syndromes are measured too rarely, idling errors accumulate on the data qubits, making the
decoder more likely to fail.  
In contrast, the more often they are measured, the more significant the faults induced by noisy measurement circuits and detectors become. %
The optimal syndrome interval therefore results from a trade-off between errors accumulated while waiting and by measuring~\cite{girvin2023introduction}.

Balancing this trade-off is especially important when quantum processors operate in a non-stationary noise environment. 
Drifts, leakage, heating, and rare correlated-noise bursts can change physical error rates during operation.
Important examples are short but strong bursts of correlated noise, which pose a major obstacle for quantum error correction~\cite{wilen2021correlated,mcewen2022resolving,mcewen2024resisting,bratrud2025measurement,google2025quantum,kurilovich2026correlated}. %
A possible way to fight dynamic changes in the noise environment is via \emph{strategic quantum error correction},  which adapts the code operations in response to information inferred from the device in vivo~\cite{tanggara2024strategic}.
Previous adaptive-QEC approaches primarily concentrated on dynamically updating decoder weights~\cite{combes2014situ,fowler2014scalable,huo2017learning,spitz2018adaptive,sivak2024optimization,remm2026experimentally,wang2023dgr,gicev2024quantum,bhardwaj2025adaptive,tan2026resilience}, physical calibrations~\cite{florjanczyk2016situ,magann2025fast,sivak2026reinforcement} or the subset of stabilizers measured~\cite{berthusen2025adaptive}. 

Here, we show that the timing of syndrome measurements provides a simple and powerful way to dramatically enhance fault-tolerant quantum memories
without additional qubits or a change of decoder class.
We propose a practical logical-noise ansatz, based on a phenomenological physical-noise model, that renders the syndrome-interval optimization problem analytically tractable.
We verify the validity of the ansatz with extensive numerical simulations on rotated surface codes with matching decoding.
With this ansatz, we show that the optimal interval is inversely proportional to 
the code distance.  
Moreover, we show that the resulting logical-error rate becomes, for large distances, exponentially better than that of distance-independent schedules. %
Interestingly, we find an intrinsic asymmetry in %
schedule %
impact: measuring too frequently increases logical errors linearly, whereas measuring too rarely amplifies them %
exponentially.
In addition, our findings %
are directly relevant for experiments: %
Fitting 
our model to the Google Willow calibration data~\cite{google2025quantum}, we show that our optimized timing predict up to $40\%$ reduction in logical error rates per unit of time.

We also %
extend our analysis to time-dependent idling noise. For (experimentally common) noise bursts, we show how to outperform every fixed-interval schedule with an adaptive protocol %
that uses long intervals during quiet periods and short intervals during bursts. %
We derive closed-form expressions to quantify 
this advantage and find that, for short but strong noise bursts, the improvement scales nearly 
linearly -- up to a small logarithmic factor-- with the relative burst amplitude. %
Finally, %
our numerical simulations not only confirm the predicted %
advantages %
but also show a nearly factor-of-two reduction
in logical failure rate for a distance-$15$ memory under time-dependent noise with realistic parameters.  
Our work establishes %
simple timing guidelines that substantially improve fault-tolerant quantum memories. %

\begin{figure}[t!]
\centering
\subfigimg[width=0.95\columnwidth]{}{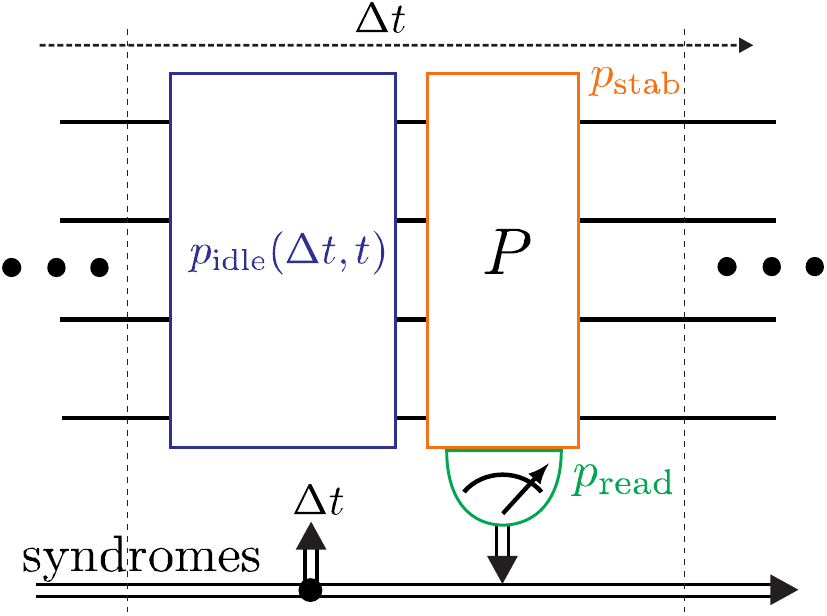}
\caption{
{\bf Optimal syndrome measurement timing in fault-tolerant quantum memories.}
For each syndrome-measurement round lasting physical time $\dt$, faults occur due to finite qubit lifetime with probability $p_\text{idle}(\dt,t)$ that increases with $\dt$. Syndrome measurements induce noise which is independent of $\dt$, where faults in data qubits occur with probability $p_\text{stab}$ and readout errors in the measuring device with probability $p_\text{read}$.
After the final syndrome measurement round, the decoder uses the space-time history of obtained syndromes to infer the best correction operation to recover the original logical state.
We choose $\dt$ to optimally trade off errors due to syndrome measurement and idling noise, 
where $\dt\propto1/d$ scaling with code distance $d$ yields exponentially lower logical error rates compared to distance-independent $\dt$.
We further reduce logical error rate for time-dependent idling noise rate $p_\text{idle}(\dt,t)$ by adaptively changing $\dt$ every round depending on syndrome measurement activity of past rounds. %
}
\label{fig:sketch}
\end{figure}

\prlsection{Phenomenological noise model}
Quantum error-correcting codes protect logical quantum states from noise by performing many rounds of syndrome measurements, where each round lasts time $\dt$. After a final round of measurements, a decoder uses the space-time history of the obtained syndromes to infer the errors that occurred and determine a recovery operation. 

During each round of syndrome measurement, faults can be broadly divided into two classes according to their origin, which is summarized in Fig.~\ref{fig:sketch}: (i) noise that continuously affects physical qubits even without our intervention, and (ii) faults induced by application of the noisy syndrome measurement. %
We model these processes within the phenomenological noise model ~\cite{dennis2002topological}:
For the first origin, even when no action is performed on a data qubit during a time $\dt$, waiting or \emph{idling} noise affects the qubit with probability
\begin{equation}
    p_{\rm idle} =1-\exp[-p\,\lambda\,\dt]\sim p\,\lambda\,\dt,
    \label{eq:pwait-full}
\end{equation}
where $p$ is the base physical noise scale, $\lambda$ is a dimensionless rate multiplier, and the right-hand side is the limit of small $p$. Such idling noise is usually dominated by the combined effect of the $T_1$ and $T_2$ times, which describe amplitude damping and dephasing noise, respectively~\cite{nielsen2010quantum,tomita2014low,krantz2019quantum}, but the explicit structure may depend on the experimental platform.
To correct errors, we perform syndrome measurements at regular time intervals $\dt$ where one measures a set of stabilizers, each with outcome $s\in\{0,1\}$. 
However, the syndrome measurements themselves are noisy, and introduce a fault on each data qubit with probability $p_\text{stab}=p$.
For the data qubits, the two independent mechanisms $p_\text{stab}$ and $p_\text{idle}$ give the total data qubit error rate %
of at least one fault occurring
\begin{equation}
    p_\text{data}=1-(1-p_\text{stab})(1-p_\text{idle})\sim p\,(1+\lambda\dt).%
    \label{eq:pd-full}
\end{equation}
Finally, the readout of the syndrome outcome $s$ itself is noisy, where its value is flipped with probability
\begin{equation}
\label{eq:read_out_error}
    p_\text{read}=b_\text{read}\,p,
\end{equation} 
with readout noise coefficient $b_\text{read}$~\cite{dennis2002topological}. The choice $b_\text{read}=0$ corresponds to the capacity noise model where syndrome readout is error-free, while $b_\text{read}>0$ captures the phenomenological noise model~\cite{dennis2002topological}. %

\prlsection{Exponential advantage from optimal intervals}
Essentially, Eq.\eqref{eq:pd-full} already encapsulates the fundamental trade-off in quantum error correction memories: When measuring too rarely, i.e. large $\dt$, data qubits suffer errors with high probability $p\,\lambda\,\dt$, which may be larger than what the code can tolerate.
In turn, measuring too excessively, i.e. small $\dt$, implies that the many noisy measurements introduce errors that degrade the logical state.
Therefore, quantum memories need to balance $\dt$ such that  logical error probability $\pL$ is as low as possible after total runtime $T$.
Thus, the key quantity is the logical error rate per physical unit of time~\cite{marton2025optimal,maurya2025synchronization,mundada2026heterogeneous}
\begin{equation}
    R=\frac{\pL}{T}.
\end{equation}
Now, we choose a sufficiently large number of syndrome rounds $T/\dt=O(d)$ to suppress time-like errors and  $\pL\ll1$ to avoid saturation effects. In this limit, $R$ becomes nearly independent of $T$ as $\pL$ scales linearly with $T/\dt$~\cite{marton2025optimal}. 
With these observations in mind, we propose the following ansatz
\begin{equation}
    R = \frac{1}{\dt}\frac{A}{d^\beta}
    \left(\frac{p}{\pth}\right)^{(d+1)/2}\left(1+\lambda \dt\right)^{g(d+1)/2},
    \label{eq:ansatz}
\end{equation}
where $A$ is a nonuniversal fitting prefactor, $\pth$ the code threshold, $\beta$ a scaling factor and
$g\in[0,1]$ describes the fraction of the leading logical exponent that is sensitive to the $\dt$-dependent idling noise.
In the limit $g=1$, one recovers %
the standard fitting formula $R\propto (p_\text{data}/\pth)^{(d+1)/2}$,
which has been widely applied in simulations~\cite{dennis2002topological,fowler2013analytic,bravyi2013simulation,bravyi2024high} and experiment~\cite{google2025quantum,bluvstein2026fault}.
We numerically verify our ansatz~\eqref{eq:ansatz} for the rotated surface code under a phenomenological bit-flip noise model decoded with PyMatching~\cite{higgott2022pymatching} (see Supplemental Material (\SM{})~\ref{sec:montecarlo}), where it works very well over a wide range of $d$, $p$, $\dt$ and $\lambda$ (see \SM{}~\ref{sec:ansatz}). 
In fact, for $b_\text{read}=1$, we usually only need to fit $A$, while the fit has stable $\beta\approx2$ and $g\approx0.8$. The code threshold $\pth(b_\text{read})$ depends on $b_\text{read}$ and is acquired from the usual threshold, where for example $\pth(b_\text{read}=1)\approx0.029$~\cite{wang2003confinement} while we determine it for other values of $b_\text{read}$ in \SM{}~\ref{sec:threshold}.

With our model, we now compute the optimal syndrome measurement interval (see \SM{}~\ref{sec:optimal})
\begin{equation}\label{eq:optimaldt}
    \dt^\star=\underset{\dt}{\text{argmin}}\, R(\dt) =\frac{2}{\lambda (g(d+1)-2)}\sim \frac{2}{\lambda gd}
\end{equation}
where the right-hand side is the limit of large $d$.
With this optimal choice, one achieves a relative reduction in logical error rate over distance-independent $\dt$
\begin{equation}
    \Gamma_{\rm opt}=\frac{R(\dt)}{R(\dt^\star)}\sim\frac{2e^{-1}}{\lambda\dt gd}(1+\lambda\dt)^{gd/2}.
    \label{eq:advantage_d}
\end{equation}
Thus, by scaling the syndrome interval inversely with distance, i.e. $\dt^\star\propto 1/d$, we gain an exponential reduction in logical error rate compared to any distance-independent interval $\dt$, which is the main result of our work.

For large $d$ and fixed $\dt/\dt^\star$, we can rewrite~\eqref{eq:advantage_d} into 
\begin{equation}
    \Gamma_{\rm opt}\sim\frac{1}{e}\frac{\dt^\star}{\dt}\exp\left({\frac{\dt}{\dt^\star}}\right).
\end{equation}
This reveals the intrinsic asymmetry in syndrome intervals: 
Measuring syndromes too often with $\dt/\dt^\star<1$ yields a linear disadvantage $\Gamma_{\rm opt}\propto\dt^\star/\dt$.  Instead, measuring too rarely, i.e. $\dt/\dt^\star>1$, gives an exponentially growing disadvantage in $\dt/\dt^\star$. 

\prlsection{Advantage of adaptive strategies}
Next, we assume that the idling noise rate $\lambda(t)$ is dynamically changing in time $t$.
A common dynamical noise model is a two-level burst process, where low noise levels $\laml$ can suddenly burst into high noise $\lamh$
\begin{equation}
\lambda(t)=\lambda_{\rm low}+(\lambda_{\rm high}-\lambda_{\rm low})\chi(t),
\,\,\,
f=\frac{1}{T}\int_0^T\chi(t)\,dt,
\label{eq:burst-env}
\end{equation}
where \(\chi(t)\in\{0,1\}\) marks high-noise periods, $f$ is the burst-time fraction, and burst noise ratios $r=\lamh/\laml$. 
Fixed time-independent intervals $\dt$ cannot react to the burst; as such one must choose a fixed $\dt$ that balances the logical error rate per unit of time $R_{\rm fixed}^\star=\min_{\dt} R_\text{fixed}(\dt)$ over both high and low noise events.
In contrast, adaptively controlling $\dt(t)$ in time allows us to react to temporal changes in $\lambda(t)$, achieving lower logical error rates $R_{\rm adapt}^\star=\min_{\dt(t)} R_\text{adapt}[\dt(t)]\leq R_{\rm fixed}^\star$.
In \SM{}~\ref{sec:adaptive_advantage}, we show that for large burst ratios $r$ and distances $d$, the relative improvement of adaptive protocols over the best fixed-time protocol is given by
\begin{equation}
    \Gamma_\text{adapt}(f^\star,r)=\frac{R_{\rm fixed}^\star(f^\star,r)}{R_{\rm adapt}^\star(f^\star,r)}\sim \frac{r}{e\log r},
    \,\,\,
    f^\star(r)\sim \frac{1}{r\log r}.
    \label{eq:large-r}
\end{equation}
where $f^\star(r)$ is the burst time fraction that exhibit the highest advantage and we assume an ideal controller with instantaneous knowledge of $\lambda(t)$.
Thus, adaptive protocols yield the best benefit for short but strong bursts, with the advantage scaling sublinearly with $r$.  For example, we find $\Gamma_\text{adapt}(r=10)\simeq1.86$ and $\Gamma_\text{adapt}(r=20)\simeq2.73$.

\prlsection{Noise-burst detection from syndrome activity}
In practice, we can obtain information about the change of $\lambda(t)$ in time from the syndromes themselves~\cite{wagner2021optimal,wagner2023learning,tan2026resilience,chadwick2024averting}: the rate of syndrome flips is directly correlated with the noise rate of the qubits. %
Based on this observation, we can detect changes in the noise level upon which we can adapt $\dt(t)$ in time. Let us assume we start with low-noise $\lambda=\laml$ and low-noise interval $\dt=\dt_\text{low}$. 
Then, we monitor the syndrome activity every round, and adjust to high-noise interval $\dt=\dt_\text{high}$ when the observed syndrome activity is consistent with a burst occurring with $\lambda=\lamh$. %
However, the syndromes are random events, with the overall syndrome activity randomly fluctuating due to shot noise.

the observed syndrome activity is consistent
with a burst, we use the detector-wise log-likelihood score~\cite{hesner2025using,tan2026resilience}
\begin{equation}
    \ell=\frac{1}{N_\text{c}}\sum_{i=1}^{N_\text{c}}
    \log\frac{P(D_{i}\mid \lamh,\dt)}
             {P(D_{i}\mid \laml,\dt)} ,
    \label{eq:llr-round}
\end{equation}
where $N_\text{c}$ is the number of stabilizer checks and $P(D_i\mid\lambda,\dt)$ is the marginal probability of detector outcome $D_i\in\{0,1\}$.  %
To further increase accuracy, we compute the moving average $\bar\ell^{(W)}$  over $W$ previous rounds of $\ell$ to reduce the chance of misclassification.
Then, whenever $\bar\ell^{(W)}\ge\theta$ with some threshold $\theta$, we predict that a burst has occurred and thus decrease $\dt$ for a pre-determined hold-time $fT$. 
The window $W$ and threshold $\theta$ balance statistical confidence against detection latency, false triggers, and missed bursts.

With an increasing number of stabilizer checks, $\ell$ becomes more
reliable due to the increasing amount of syndrome information.
Although neighboring detector outcomes are locally correlated, under
our phenomenological noise model they form a dependency graph of
bounded degree.  Consequently, the $W N_c=\Theta(Wd^2)$ detector
samples in a $W$-round window obey an exponential concentration
bound, with the
false-positive and false-negative probabilities satisfying (see \SM{}~\ref{sec:finite_window_detection})
\begin{equation}
    P_{\rm FP},P_{\rm FN}
    =
    \exp(-\Omega(Wd^2)).
    \label{eq:detection_concentration}
\end{equation}
Thus, for sufficiently large codes, even a small number of syndrome
rounds can identify a change in the noise level with high confidence.

\begin{figure}[t]
    \centering
    \subfigimg[width=0.24\textwidth]{a}{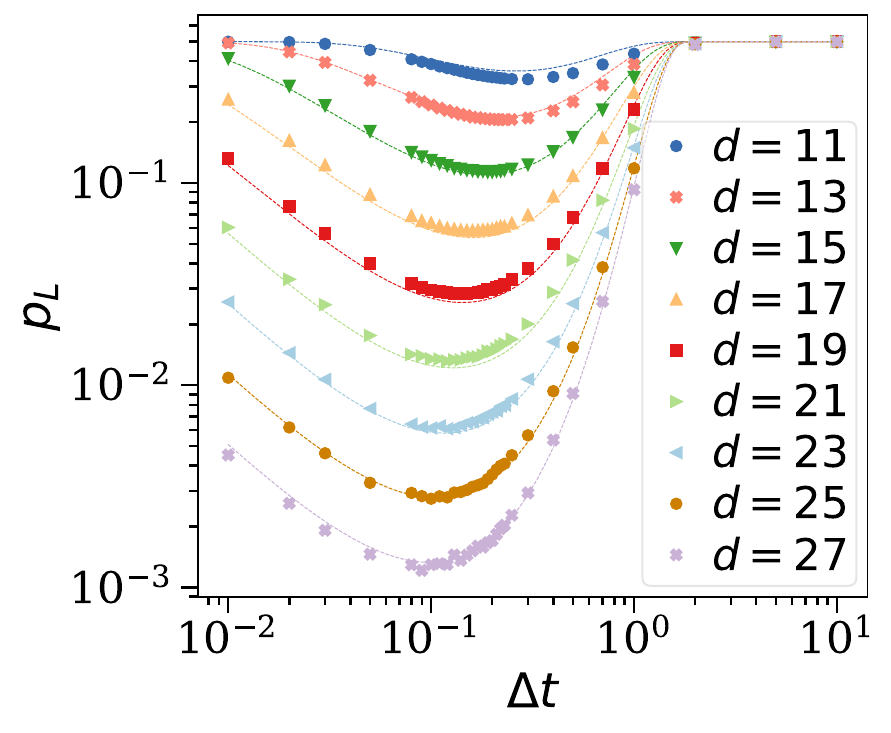}\hfill
        \subfigimg[width=0.24\textwidth]{b}{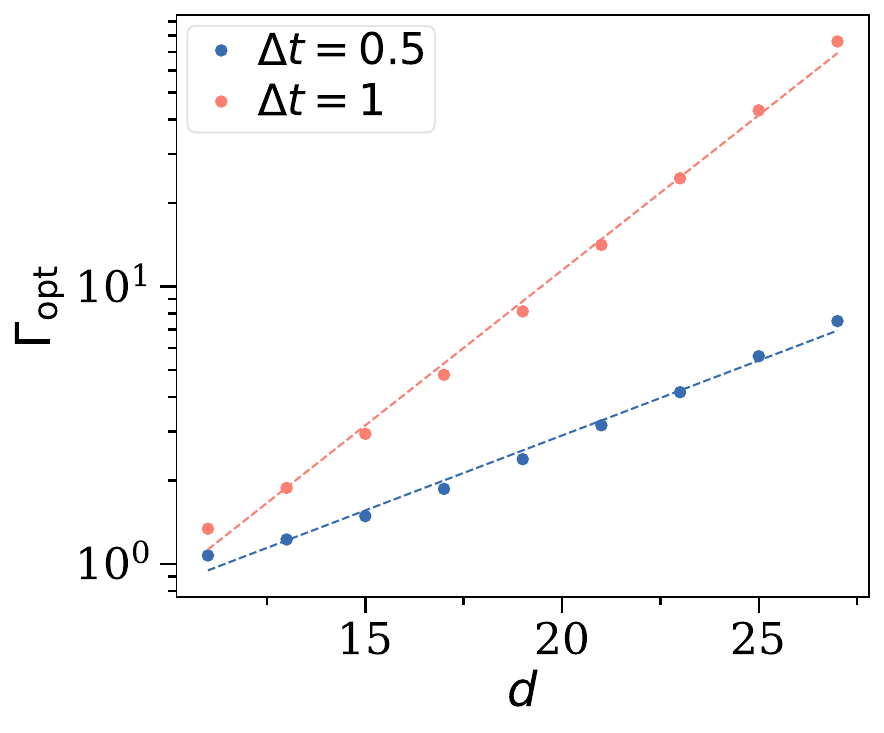}

	\setbox1=\hbox{
    \begin{tabular}{c|cccccccccccc}
\hline
\(d\) & 3 & 5 & 7 & 9 & 11 & 13 & 17 & 19 & 23 & 27 \\
\hline
\hline
\(\dt^\star/\dt\)
& 5.38 & 2.30 & 1.47 & 1.08 & 0.85 & 0.70  & 0.52 & 0.46 & 0.38 & 0.32 \\
\hline
\(\Gamma_{\rm opt}\)
& 1.72 & 1.21 & 1.05 & 1.00 & 1.01 & 1.06 & 1.24 & 1.37 & 1.74 & 2.28 \\
\hline
\end{tabular}
    }%
	\leavevmode\rlap{\usebox1}%
	\rlap{\hspace*{0pt}\raisebox{\dimexpr\ht1+0.2\baselineskip}{{\bfseries \large\textsf{c}}}}%
	\phantom{\usebox1}%
    \caption{
    {\bf Quantum memory enhancement by optimal syndrome-measurement timing for static noise}. We simulate the rotated surface code for different distances $d$ under a phenomenological noise model with time-independent idle rate $\lambda=1$, base noise scale $p=0.015$, readout error factor $b_\text{read}=1$ and total time $T=200$. 
    \idg{a} We show the total logical error $\pL$ obtained from the full simulation %
    (dots) and from ansatz~\eqref{eq:ansatz} (dashed lines) where we fitted $A=1.7$ while parameters $\beta=2$, $g=0.8$ and $\pth=0.029$ are fixed. %
    \idg{b} Advantage factor $\Gamma_\text{opt}$ in $\pL$ at the optimal choice of distance-dependent syndrome interval $\dt^\star$ over fixed interval $\dt$, for two different $\dt$'s. Dashed lines are 
    fits with $\Gamma_\text{opt}\propto \exp(\gamma d)$, for %
    $\gamma=\{0.12,0.25\}$.
    \idg{c} Estimated $\Gamma_\text{opt}$ and $\dt^\star/\dt$ for different $d$ for the Google Willow experimental calibration data~\cite{google2025quantum} with experimental $\dt=1.108\mu s$ as constant baseline (see \SM{}~\ref{sec:supp_google_mapping}). %
    }
    \label{fig:fixed}
\end{figure}

\prlsection{Numerical results}
We %
numerically validate our findings 
for rotated surface-code memories under the phenomenological noise model. 
First, in Fig.~\ref{fig:fixed}, we study time-independent idle noise at constant %
$\lambda$. We show the %
logical-error probability %
$\pL$ after time $T$ against $\dt$ for different code distances $d$ in Fig.~\ref{fig:fixed}a. We find that the full simulation (dots) closely matches the ansatz~\eqref{eq:ansatz} (curves%
) which we fit with parameter $A$. We observe that the optimal interval scales as $\dt^\star \propto 1/d$, which we confirm via a fit in \SM{}~\ref{sec:fixedt}. In Fig.~\ref{fig:fixed}b, we show that the improvement $\Gamma_\text{opt}$ in logical error of $\dt^\star$ over distance-independent $\dt$ scales exponentially in $d$.
In Fig.~\ref{fig:fixed}c, we show the predicted $\Gamma_\text{opt}$ and $\dt^\star$ for different distances $d$ for the recent Google Willow experiments of Ref.~\cite{google2025quantum}. For this, we fit the %
experimental calibration data of Ref.~\cite{google2025quantum} onto our phenomenological model (see \SM{}~\ref{sec:supp_google_mapping}).
According to our model, %
the %
experimental interval is shorter than the predicted optimum for $d=3,5,7$, where larger $\dt$ could yield up to a $40\%$ reduction in logical error rate per unit %
time. In contrast, we predict that potential $d>9$ experiments could improve via smaller $\dt$ than implemented in Ref.~\cite{google2025quantum}, assuming  syndrome measurements can be realized in shorter time $\dt$ with constant physical error rates.

\begin{figure}[t]
    \centering
    \subfigimg[width=0.49\textwidth]{}{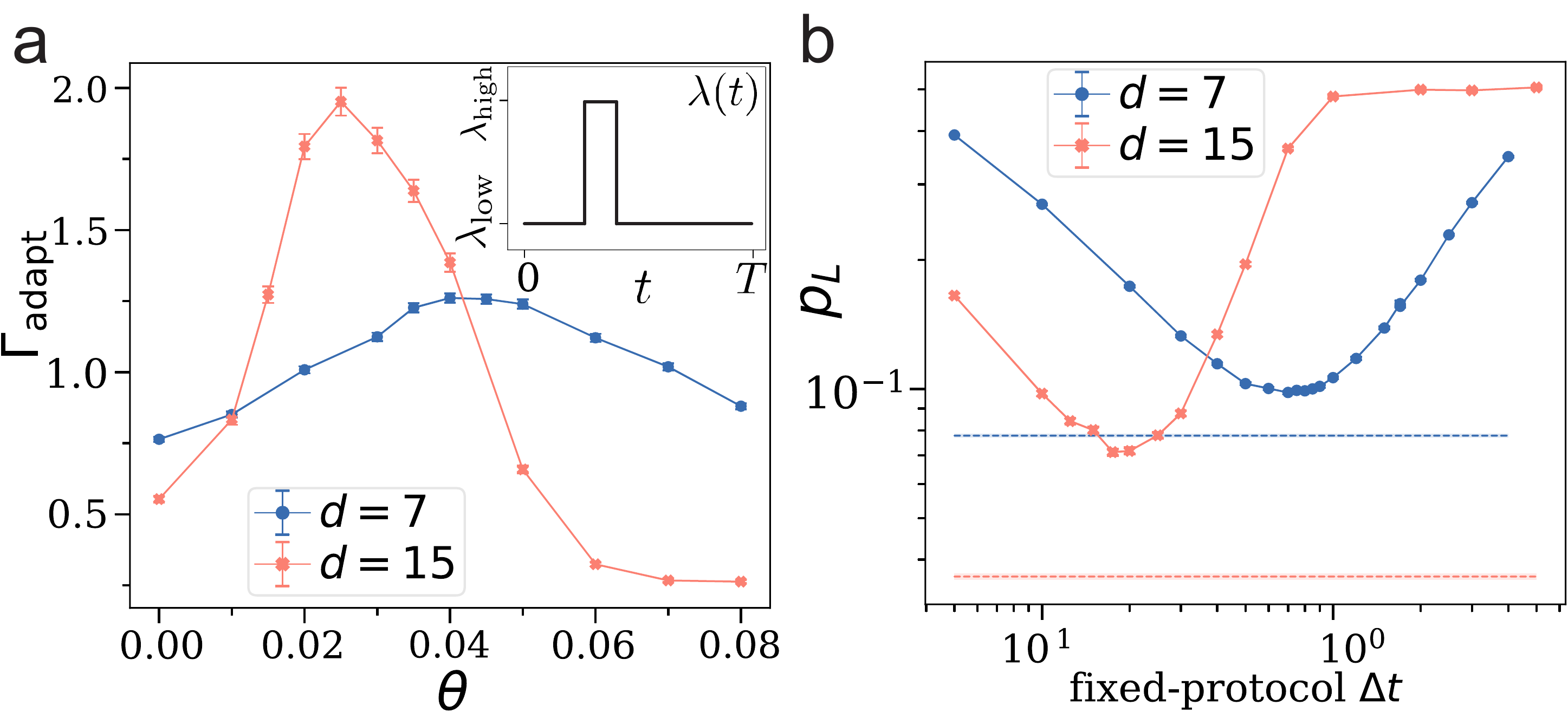}
    \caption{%
    {\bf Adaptive syndrome measurement timing for dynamic noise.} We consider burst noise~\eqref{eq:burst-env} in a rotated surface code with distance $d$.
    \idg{a} Improvement in logical error  $\Gamma_\text{adapt}$ of adaptive protocols over the best fixed-time protocol, plotted against burst detection threshold $\theta$. We find maximal 
    $\Gamma_\text{adapt}\approx \{1.26,1.95\}$  with $\dt^\star=\{0.7,0.175\}$. Inset shows typical burst idle noise rate $\lambda(t)$ over time $t$.
    \idg{b} Total logical error $\pL$ of fixed-interval protocol against fixed-protocol time $\dt$. 
    Horizontal dashed lines indicate best adaptive protocol with $\theta^\star=\{0.045,0.025\}$. 
    As parameters, for $d=7$ we have total time $T=2000$, $p=0.015$, $\laml=0.03$, $\lamh=2$, syndrome interval $\dt_\text{high}=0.25$ when burst is detected, syndrome interval $\dt_\text{low}=1.7$ outside burst, burst-time fraction $f=0.064$ with a single burst, and syndrome activity window $W=7$. 
     For $d=15$, we have $T=8000$, $p=0.01$, $\laml=0.03$, $\lamh=3$, $\dt_\text{low}=0.57$, $\dt_\text{high}=0.061$, $f=0.029$, $W=5$.
    }
    \label{fig:adaptive}
\end{figure}

Next, in Fig.~\ref{fig:adaptive}, we study fixed-interval and adaptive-interval protocols for time-dependent burst noise $\lambda(t)$ of~\eqref{eq:burst-env}.
In Fig.~\ref{fig:adaptive}a, we show the relative advantage $\Gamma_\text{adapt}$ in logical error of our adaptive protocol over the best fixed-interval protocol as function of burst activity threshold $\theta$. For the optimal choice $\theta^\star$, we find $\Gamma_\text{adapt}\approx \{1.26,1.95\}$ for $d=\{7,15\}$. 
In Fig.~\ref{fig:adaptive}b, we show $\pL$ against fixed interval $\dt$ for two different quantum memory experiments with distance $d=7$ and $d=15$, respectively. 
We find that any fixed $\dt$ has higher logical error rate than our best adaptive protocol that adjusts $\dt$ according to the observed syndrome activity.

\prlsection{Discussion}
In this work, we significantly enhance noise suppression in fault-tolerant quantum memories by optimizing syndrome measurement intervals $\dt$. 
The choice ${\dt^\star\propto 1/d}$ optimally balances between syndrome-measurement errors %
and idling noise, 
yielding an exponential-in-$d$ reduction in logical error rates compared to any strategy with distance-independent intervals. %

We note that the measurement operations themselves are not arbitrarily fast, but require a finite time $\dt_\text{min}$, depending on the physical constraints of the experimental platform.
As such, beyond a certain $d$, the optimal scaling $\dt^\star(d)\propto 1/d$ cannot be implemented in practice, effectively limiting us to $\dt=\text{max}(\dt^\star,\dt_\text{min})$. 
Nonetheless, for finite-size codes, our formula gives a direct guidance on how to choose $\dt$ to preserve the quantum memory as long as possible.
The time needed to implement syndrome measurements $\dt_\text{min}$ can be shortened, but this usually goes hand in hand with increased readout errors~\cite{pattison2021improved}. As such, one has to trade-off $\dt_\text{min}$ between enhanced readout noise and the noise reduction from optimal syndrome timing. This trade-off could be improved via decoders that use soft information from the readout~\cite{pattison2021improved}.

As a concrete application, we map the calibration data of the recent Google Willow experiments of Ref.~\cite{google2025quantum} onto our phenomenological model. 
For their experimentally demonstrated distances, larger syndrome intervals could yield up to a $40\%$ reduction in logical error rate per unit of time (see \SM{}~\ref{sec:supp_google_mapping}).
Further, for larger distances, we predict that the optimized intervals enhance the noise suppression factor $\Lambda=\pL(d)/\pL(d+2)$, which is the decrease in logical error when increasing the distance by $2$, by $24\%$. Here, we neglect polynomial corrections in the logical error ansatz and we assume that syndrome measurements can be implemented within time $\dt^\star$, which is faster than the current experiment for $d>9$ (see \SM{}~\ref{sec:dist_reduction}).
Other platforms such as Rydberg atoms and ion traps face the same trade-off~\cite{wang2021single}, where we expect similar potential for noise reduction~\cite{bluvstein2026fault,paetznick2026improved}.

While we considered the phenomenological noise model, where the syndrome measurement errors are abstract, our approach equivalently applies to circuit-level noise models where the code operations are explicitly modeled as noisy gates. Here, the explicit form of the idling noise model depends on the experimental platform, which is usually the combined effect of amplitude damping and dephasing noise characterized by $T_1$ times and $T_2$, respectively (see \SM{}~\ref{sec:supp_google_pmem_check}).

For time-dependent noise, we gain further improvement in noise suppression by adaptively changing $\dt$ according to the syndrome activity: When the noise and thus syndrome flip activity is low, one can wait longer between subsequent syndrome measurements, to reduce the impact of measurement-induced noise. In contrast, in high-noise environments heralded by high syndrome activity, it is better to measure more often to prevent the accumulation of noise in the data qubits. 
For short but strong bursts, we analytically find 
a reduction in logical error rate that scales nearly linearly with the burst-noise ratio $r$ %
compared to any fixed-interval protocol. %
This adaptive syndrome timing protocol could help in experiments, such as in superconducting quantum computers where noise bursts are often observed due to high-energy events~\cite{wilen2021correlated,mcewen2022resolving,bratrud2025measurement,google2025quantum,kurilovich2026correlated}.
For example, such short but strong fluctuations in dephasing noise have been observed in superconducting qubit experiments, exhibiting up to one order of magnitude in relative burst strength~\cite{berritta2026real}.

While the syndrome information itself is noisy and subject to shot-noise, we nonetheless can identify high-noise events by reacting to events that exceed a threshold in the log-likelihood of the syndrome activity. With this, the rotated surface code under a phenomenological burst noise model and PyMatching decoding yields nearly a factor-of-two improvement for $d=15$.

More broadly, our protocol provides an example of strategic QEC, in which syndrome information controls the code operation itself rather than only the decoder~\cite{tanggara2024strategic}. Extensions could adapt timing or code structure to adversarial and spatiotemporally correlated noise~\cite{arvind2025quantum,amezcua2026assessing,stephens2008asymmetric,hill2011fault}.

\prlsection{Acknowledgments}
K.B. is supported by a Hartree fellowship from the Joint Center for Quantum Information and Computer Science (QuICS) at the University of Maryland, College Park.
ChatGPT 5.5 was used to assist with language, clarity, proving and coding. All suggested changes were reviewed and edited by the authors, who take full responsibility for the final content.

\prlsection{Data availability}
The data cannot be made publicly available upon publication because they are not available in a format that is sufficiently accessible or reusable by other researchers. The data that support the findings of this study are available upon reasonable request from the authors.

\bibliography{refs}

\let\addcontentsline\oldaddcontentsline

\appendix
\onecolumngrid
\newpage

\setcounter{secnumdepth}{2}

\renewcommand{\thesection}{\Alph{section}}
\renewcommand{\thesubsection}{\arabic{subsection}}

\begin{center}
{\large \textbf{Supplemental Material}}
\end{center}

In the Supplemental Material, we provide further proofs underlying our main results as well as additional findings.

\tableofcontents

\section{Phenomenological noise simulation of rotated surface code}\label{sec:montecarlo}

In the main text, we simulate a rotated surface-code memory experiment against 
bit-flip noise. We sample phenomenological data faults during the experiment waiting intervals, $Z$ syndrome-measurement faults on each round, and a final noisy readout.  The complete detector history is decoded with minimum-weight matching~\cite{higgott2022pymatching}, where we use the standard edge weights derived from the physical error model, where for the adaptive model we use the believed noise state of the controller.

\begin{figure}[htpb]
\centering
\subfigimg[width=0.4\columnwidth]{}{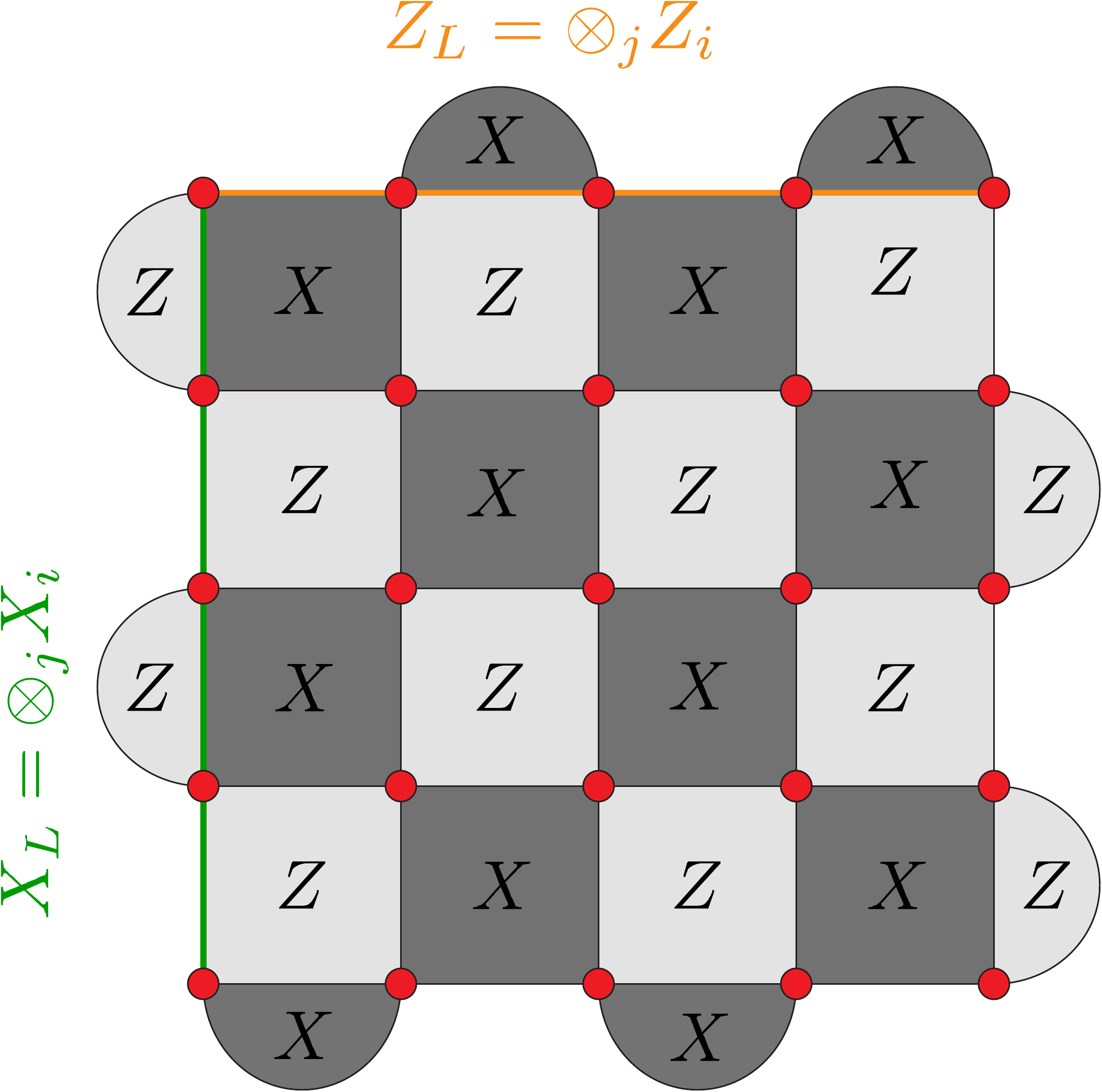}
\caption{Rotated surface code for $d=5$. 
}
\label{fig:surfacecode}
\end{figure}

We use the odd-distance rotated surface code, which is a planar CSS stabilizer code on a $d\times d$ array of data qubits (see Fig.~\ref{fig:surfacecode}).  It has one encoded qubit and distance $d$.  Stabilizers are arranged on alternating plaquettes: one sublattice contains $X$ checks and the other contains $Z$ checks.  Bulk checks have weight four, while checks cut by a boundary have reduced weight.  The two boundary types are rough and smooth.  A logical operator is a string of Pauli operators connecting the corresponding pair of opposite boundaries, and the shortest such string has length $d$.

Because the code is CSS, the two Pauli sectors can be treated separately in the phenomenological model.  A $Z$-check syndrome detects $X$-type data faults, while an $X$-check syndrome detects $Z$-type data faults.  Our simulations decode only the $Z$-check  sector; the other sector is equivalent after exchanging $X$ and $Z$.

The decoder input is a detector history.  Let $s_{j,c}$ be the measured value of check $c$ in syndrome round $j$.  The detector associated with check $c$ between rounds $j-1$ and $j$ is
\begin{equation}
    D_{j,c}=s_{j,c}\oplus s_{j-1,c}.
    \label{eq:S-detector}
\end{equation}
A data fault on a qubit flips the parity of the adjacent checks, and therefore creates a spatial detector pattern at the corresponding time layer.  A syndrome-bit fault flips the reported outcome of one check in one round, and therefore creates a pair of detectors separated in the time direction.  Boundary data faults and terminal readout faults create detector patterns connected to boundary nodes.  The full history is therefore represented by a three-dimensional matching graph whose spatial directions are the code patch and whose time direction is the sequence of syndrome rounds.

The Monte Carlo simulation of the code is performed as follows:  First, for time-dependent $\lambda(t)$, we choose the next syndrome interval $\dt_j$ depending on our burst detection protocol of \SM{}~\ref{sec:finite_window_detection}. For time-independent $\lambda=\text{const}$, we keep a fixed $\dt$.
Second, the true environment $\lambda(t_j)$ determines the data-fault probability through Eq.~\eqref{eq:pd-full}.  Data faults and measurement faults are sampled independently.  Third, the syndrome bits are updated and converted into detectors through Eq.~\eqref{eq:S-detector}.  This process is repeated until the accumulated physical time reaches $T=\sum_j \dt_j$, where we measure a final noisy readout of all data qubits and append the outcome to the detector history.

Decoding is performed by minimum-weight perfect matching on the complete detector graph~\cite{higgott2022pymatching}.  The edge weights are log-likelihood weights derived from the decoder-believed probabilities,
\begin{equation}
    w_e=\log\frac{1-p_e}{p_e},
    \label{eq:S-edge-weight}
\end{equation}
with distinct weights for space-like data edges and time-like measurement edges.  
For fixed-interval simulations we assume that the decoder knows the true instantaneous noise rate $\lambda(t)$ for every $t$. 
For adaptive detector-threshold controllers the decoder belief follows the controller state, not the hidden true noise rate $\lambda(t)$.  A shot fails if the correction returned by matching differs from the accumulated physical error by a nontrivial logical operator.

For the analytic fits according to~\eqref{eq:ansatz_sup}, we map the analytic logical error rates per unit of time to logical errors over total time $T=\sum_j \dt_j$ via
\begin{equation}
    \pL=\frac{1}{2}(1-\exp(-2\sum_j R(\dt_j,t_j)\dt_j))
\end{equation}
where we sum over all rounds $j$, where time $t_j=\sum_{i=1}^{j-1}\dt_i$. One can verify that for $\sum_{j}R(\dt_j,t_j)\dt_j\gg1$, we have $\pL=\frac{1}{2}$ as expected for a completely random binary variable (which here corresponds to the outcome of the logical Pauli Z operator), while for $\sum_{j}^{}R(\dt_j,t_j)\dt_j\ll1$, the logical error is $\pL\approx \sum_j R(\dt_j,t_j)\dt_j$ as expected.

\section{Numerical verification of the ansatz}\label{sec:ansatz}
\begin{figure}[htpb]
\centering
\subfigimg[width=0.3\columnwidth]{a}{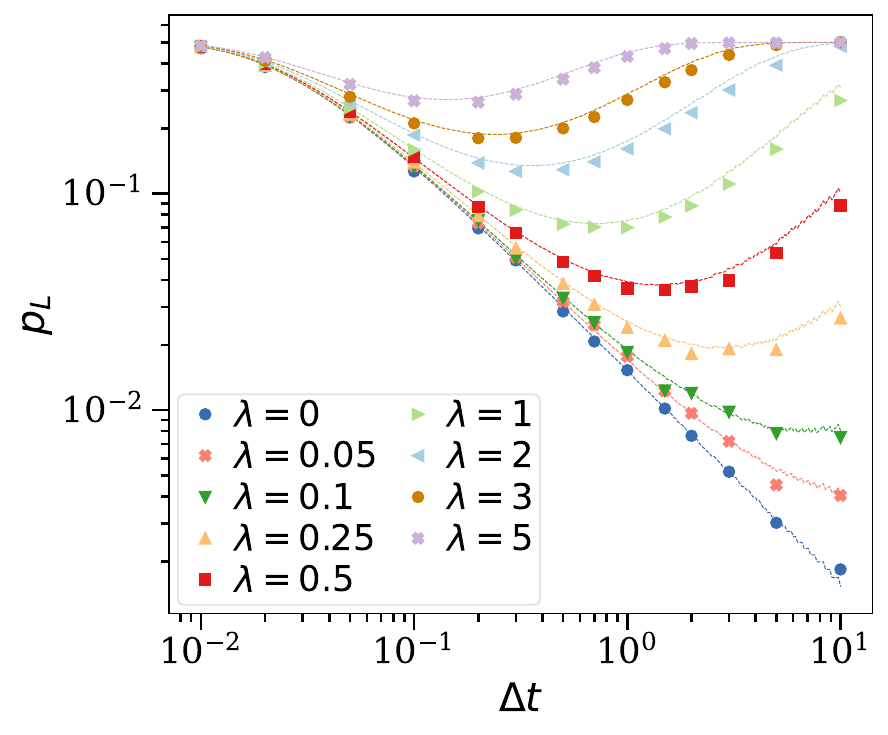}
\subfigimg[width=0.3\columnwidth]{b}{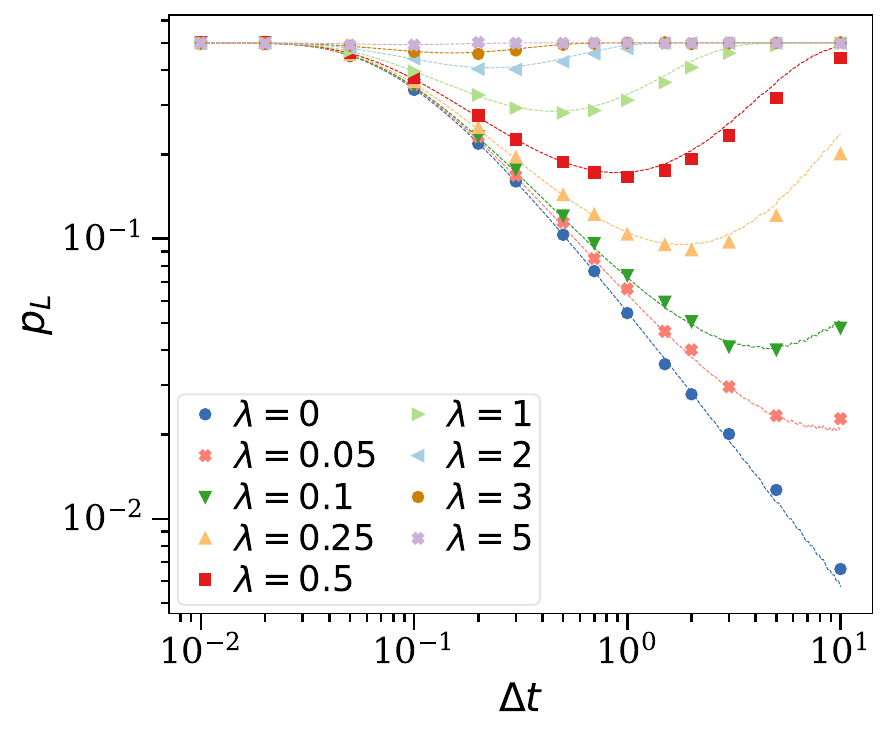}
\subfigimg[width=0.3\columnwidth]{c}{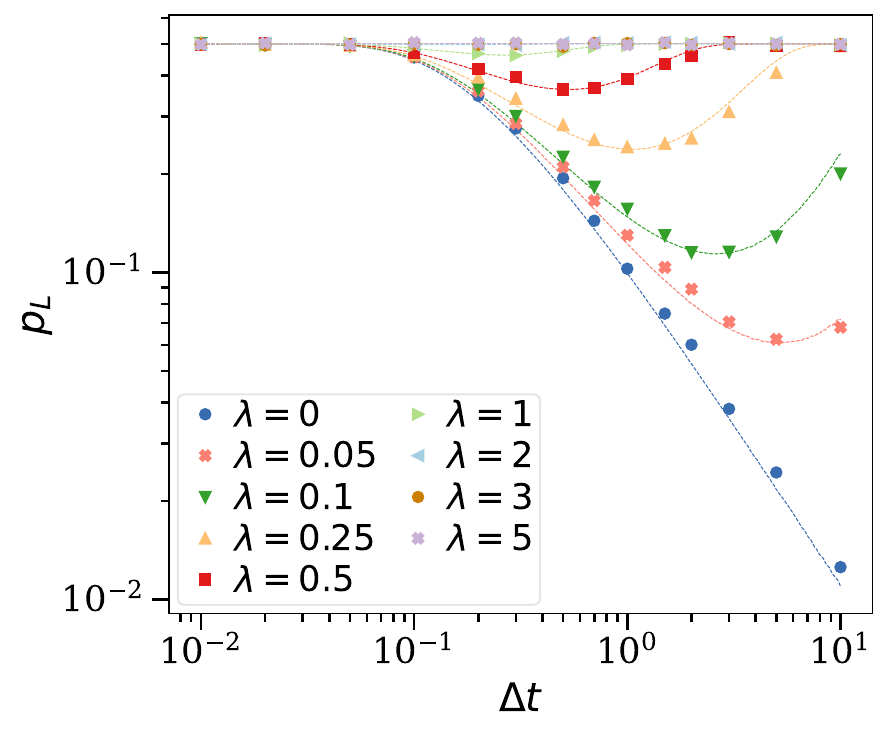}
\subfigimg[width=0.3\columnwidth]{d}{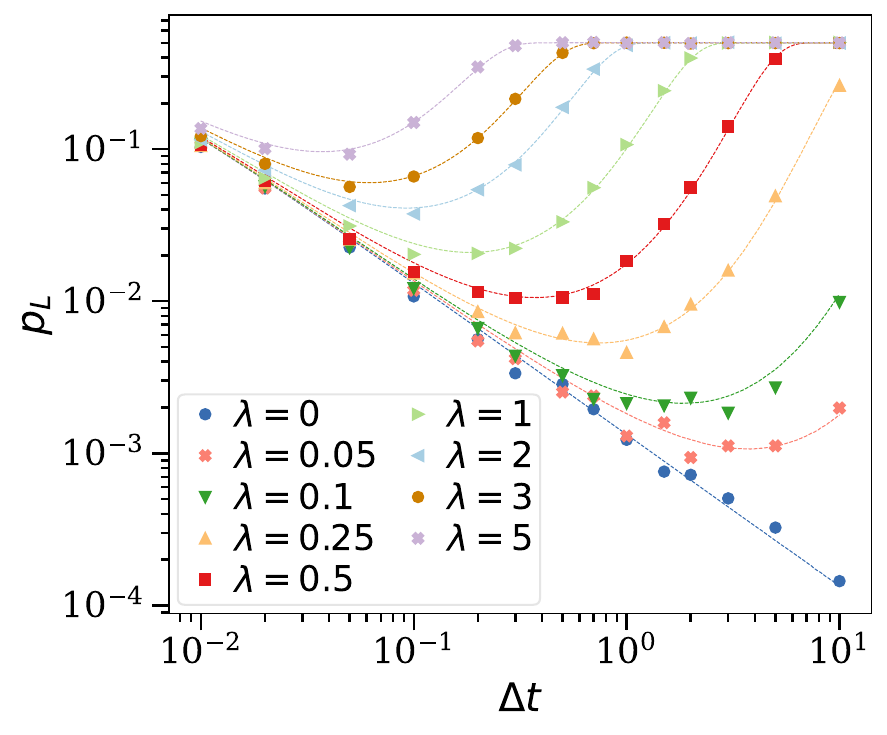}
\subfigimg[width=0.3\columnwidth]{e}{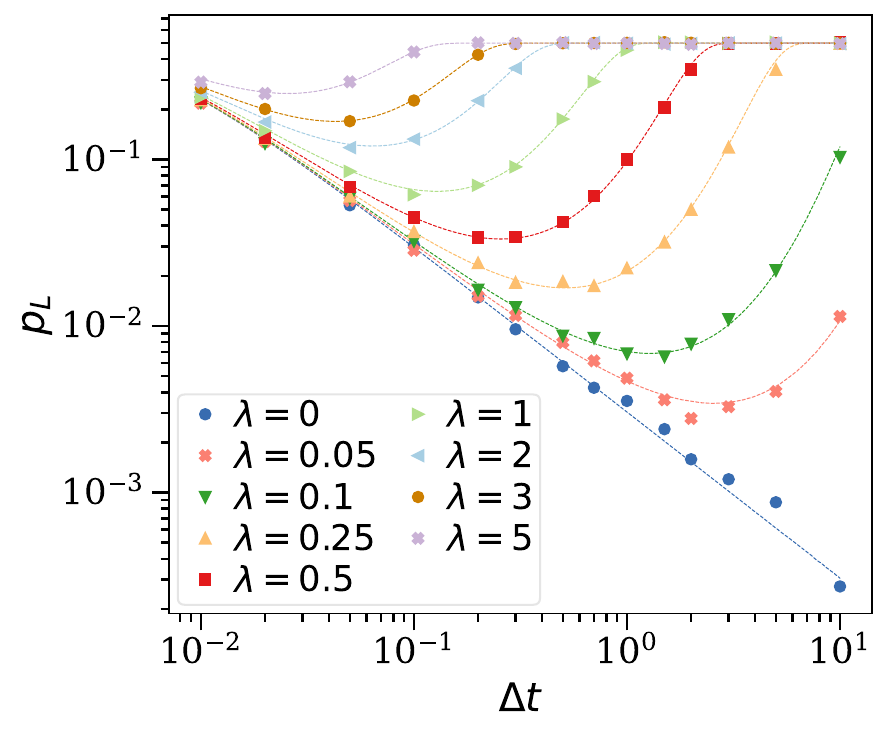}
\caption{Total logical error rate $\pL$ after time $T$ against syndrome interval $\dt$ for different constant idling noise $\lambda$. We show numerical simulation of phenomenological noise model of the rotated surface code under matching decoding (dots) and fitted ansatz of~\eqref{eq:ansatz}. We always fix ansatz parameters $\beta=2$, $g=0.8$ and $\pth=0.029$, while we fit $A$ depending on $p$ and $d$.  
For the plots, we show different distances $d$ with
\idg{a} $d=5$, $p=0.005$, $T=100$ and $A=0.75$.
\idg{b} $d=7$, $p=0.01$, $T=200$ and $A=1.0$.
\idg{c} $d=11$, $p=0.015$, $T=500$ and $A=1.4$.
\idg{d} $d=15$, $p=0.01$, $T=1000$ and $A=1.5$.
\idg{e} $d=21$, $p=0.015$, $T=1000$ and $A=1.9$. 
For all plots, we choose readout noise coefficient $b_\text{read}=1$.
}
\label{fig:ansatz}
\end{figure}

In this section, we verify the accuracy of our logical error rate ansatz from the main text, reproduced in~\eqref{eq:ansatz_sup}.
To verify, we compare the ansatz with full numerical simulations of the phenomenological noise model with the rotated surface code under matching decoding in Fig.~\ref{fig:ansatz}.
For different $\lambda$ and $d$, we find that our ansatz closely matches the full simulation. Notably, we only need to fit the $A$ parameter, while for all fits we keep fixed $\beta=2$, $g=0.8$ and $\pth(b_\text{read}=1)=0.029$.

\begin{figure}[htpb]
\centering
\subfigimg[width=0.5\columnwidth]{}{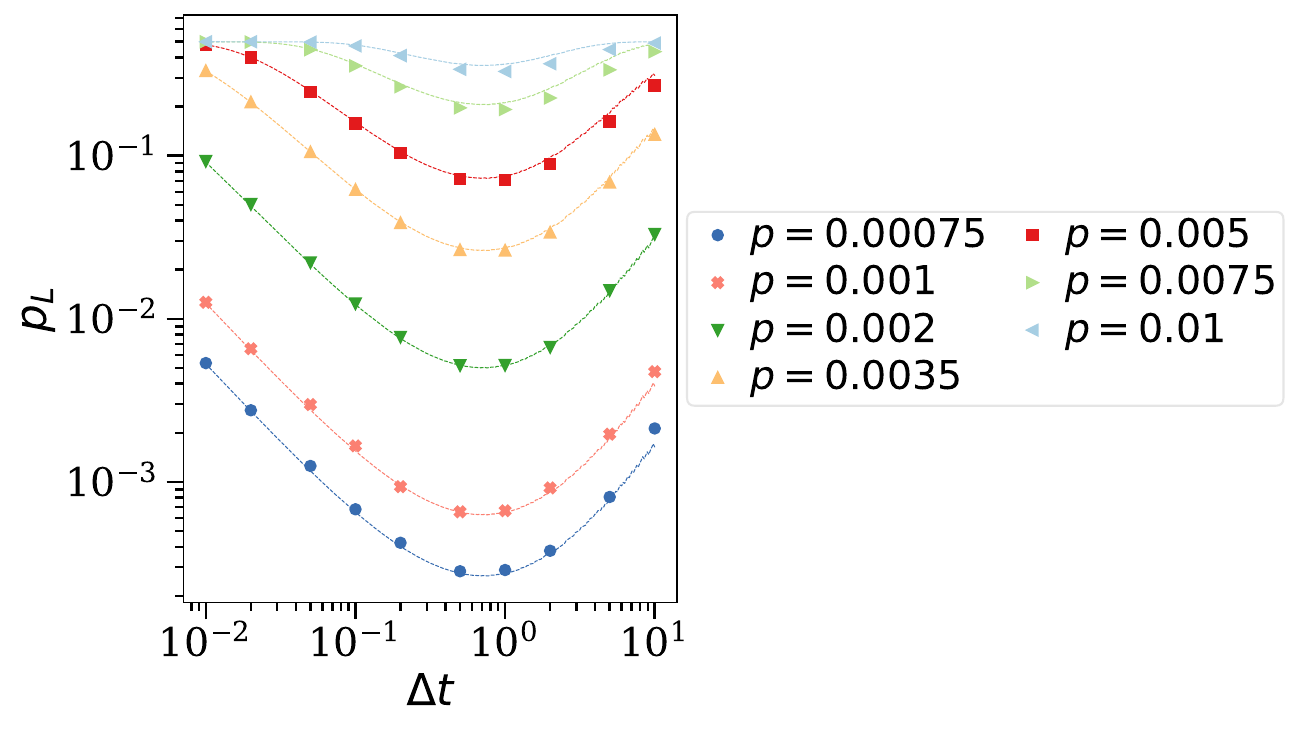}
\caption{Total logical error rate $\pL$ after time $T$ against syndrome interval $\dt$ for different physical error rates $p$. We show numerical simulation of phenomenological noise model of the rotated surface code with $d=5$ and $b_\text{read}=1$ under matching decoding (dots) and fitted ansatz of~\eqref{eq:ansatz}. We fix $\beta=2$, $g=0.8$ and $\pth=0.029$, while we fit $A=0.75$.
}
\label{fig:ansatz_p}
\end{figure}

Next, in Fig.~\ref{fig:ansatz_p}, we study our ansatz for different $p$, where we compare the rotated surface code simulation against our ansatz and fix $b_\text{read}=1$. We find again that our ansatz matches the full simulation very well, with only a single fitting parameter $A$ required.

\begin{figure}[htpb]
\centering
\subfigimg[width=0.3\columnwidth]{a}{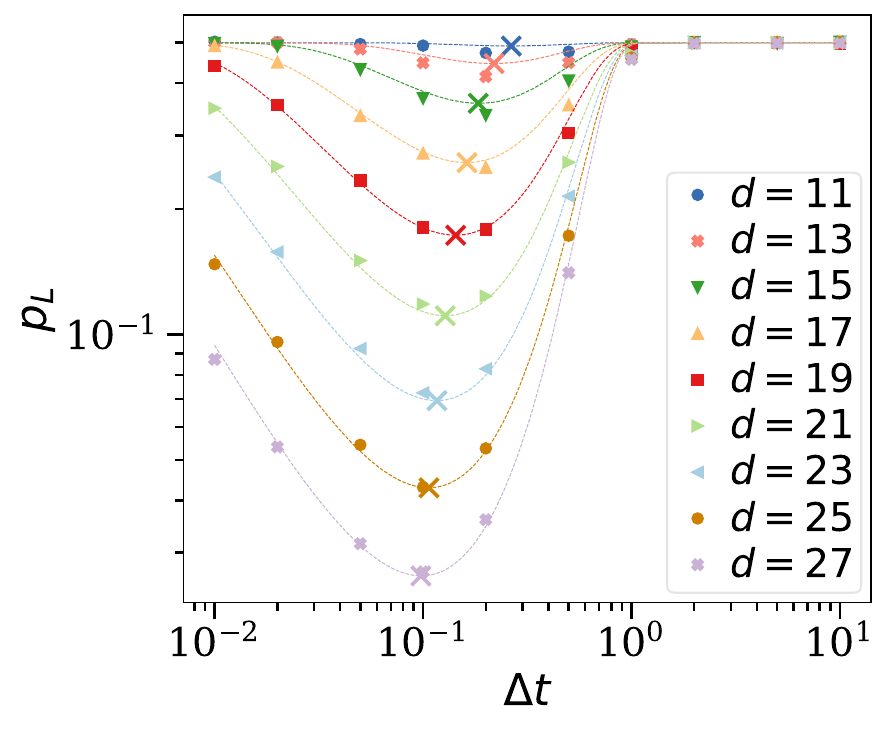}
\subfigimg[width=0.3\columnwidth]{b}{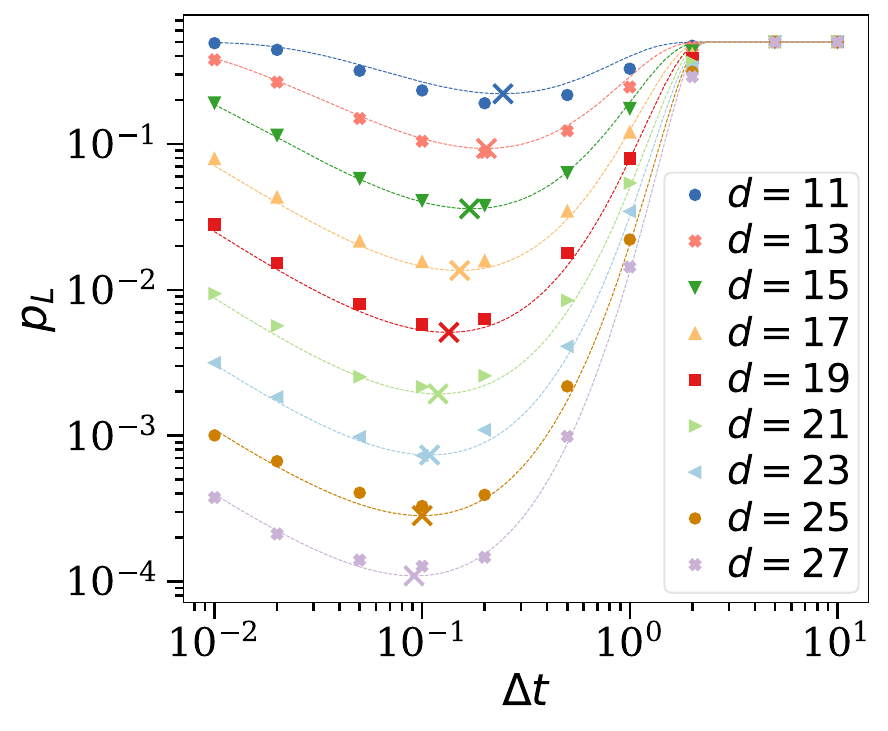}
\caption{Total logical error rate $\pL$ after time $T$ against syndrome interval $\dt$ for distances $d$ and readout measurement coefficient $b_\text{read}$. We show numerical simulation of phenomenological noise model of the rotated surface code with $p=0.015$ under matching decoding (dots) and fitted ansatz of~\eqref{eq:ansatz}. We have $\beta=2$ as usual.
\idg{a} High readout noise with $b_\text{read}=2$, where we fit $g=0.8$ and $A=1.35$, and determined $\pth(b_\text{read}=2)\approx0.023$ as the error threshold from \SM{}~\ref{sec:threshold}.
\idg{b} Low readout noise with $b_\text{read}=0.5$, where we fit $g=0.85$ and $A=2.7$. We find the error threshold $\pth(b_\text{read}=0.5)\approx0.036$ from \SM{}~\ref{sec:threshold}.
}
\label{fig:ansatz_b}
\end{figure}

Finally, in Fig.~\ref{fig:ansatz_b} we study our ansatz for different $d$ and readout measurement error coefficient $b_\text{read}=2$ (Fig.~\ref{fig:ansatz_b}a) and $b_\text{read}=0.5$ (Fig.~\ref{fig:ansatz_b}b).  We find that our ansatz matches the full simulation very well. For the fits, we use the $\pth(b_\text{read})$ as we computed in \SM{}~\ref{sec:threshold}.
For $b_\text{read}=0.5$, we fit the factor $g=0.85$. Indeed, $g$ is expected to increase with decreasing $b_\text{read}$, as for code capacity noise model, which has $b_\text{read}=0$, we have $g=1$.

\section{Numerical fitting of the code threshold}\label{sec:threshold}
\begin{figure}[htpb]
\centering
\subfigimg[width=0.32\columnwidth]{a}{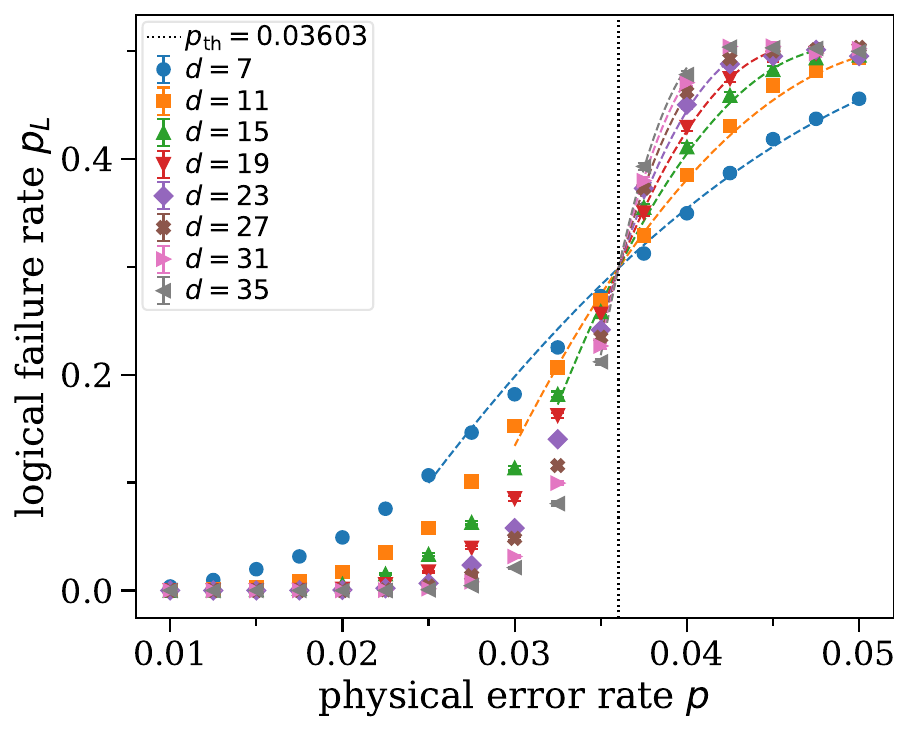}
\subfigimg[width=0.32\columnwidth]{b}{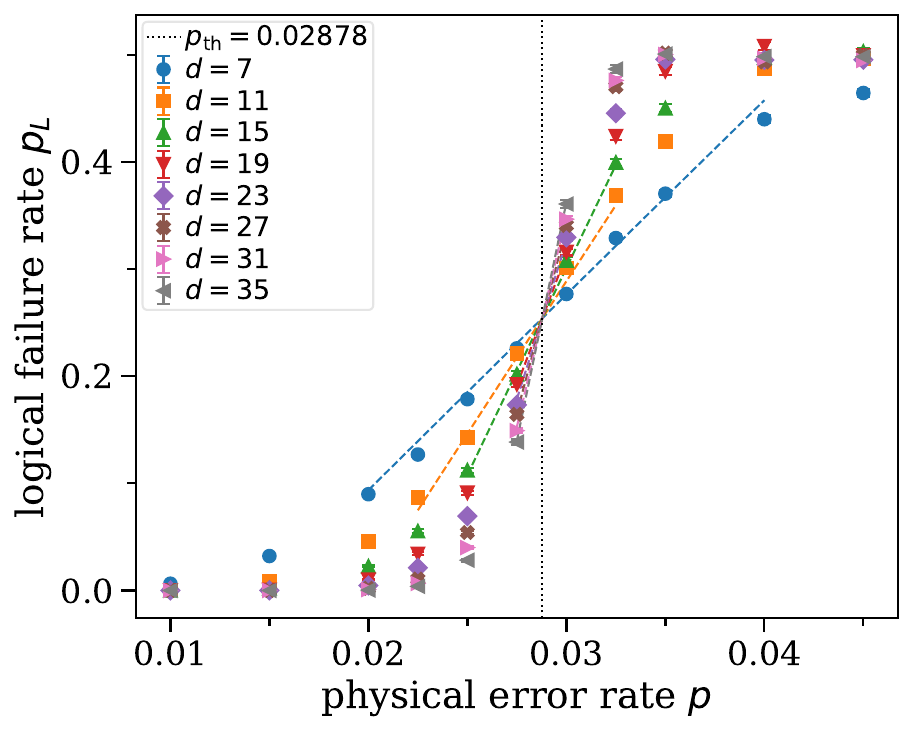}
\subfigimg[width=0.32\columnwidth]{c}{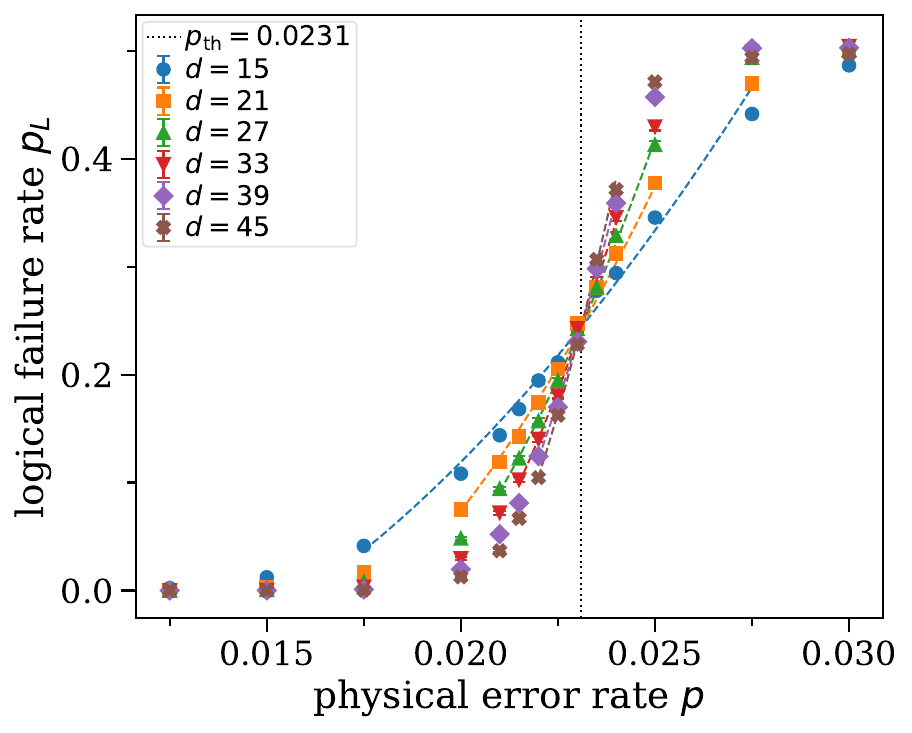}
\caption{Fit of code threshold for different readout noise coefficient $b_\text{read}$. We show logical error rate $\pL$ against $p$ after $3d$ rounds of syndrome measurements for $\lambda=0$ for different $d$. We  simulate of phenomenological noise model of the rotated surface code under matching decoding. 
For the plots, we show different distances $d$ where fitted $\pth$ is indicated as vertical dashed line, while dashed curves are polynomial fits around threshold
\idg{a} $b_\text{read}=0.5$ with fitted $\pth\approx 0.036$,
\idg{b} $b_\text{read}=1$ with fitted $\pth\approx 0.029$, and
\idg{c} $b_\text{read}=2$ with fitted $\pth\approx 0.023$.
}
\label{fig:threshold}
\end{figure}

Here, we numerically fit the noise threshold $\pth$ for the phenomenological noise model for different $b_\text{read}$ in Fig.~\ref{fig:threshold}.
We set $\lambda=0$ and run the code over $3d$ rounds of syndrome measurements, and calculate $\pL$. Then, we plot $\pL$ against $p$ for different $d$, and observe that the curves cross at a single point, which is $\pth$. To accurately determine $\pth$, we fit around the threshold $x=(p-\pth)d^{1/\nu}$ with a polynomial second-order ansatz $\pL=a_0 + a_1 x + a_2 x^2$, which is shown as dashed lines.
We find $\pth(b_\text{read}=0.5)\approx 0.036$,  $\pth(b_\text{read}=1)\approx 0.029$ and $\pth(b_\text{read}=2)\approx 0.023$.

\section{Optimal syndrome intervals}\label{sec:optimal}
We assume that the logical error rate per unit time is modeled with our ansatz from the main text
\begin{equation}
    R=\frac{1}{\dt}\frac{A}{d^\beta}
    \left(\frac{p}{\pth}\right)^{(d+1)/2}\left(1+\lambda \dt\right)^{\alpha},
    \label{eq:ansatz_sup}
\end{equation}
where $A$ is a nonuniversal fitting prefactor, $\beta$ a scaling factor and
\begin{equation}
    \alpha=g\frac{d+1}{2},
\end{equation}
where $g\in[0,1]$ describes fraction of the leading logical exponent that is sensitive to the $\dt$-dependent idling noise.
When we set $g=1$, we recover the well known fitting formula $R\propto (p/\pth)^{(d+1)/2}$~\cite{dennis2002topological,fowler2013analytic,bravyi2013simulation,bravyi2024high}.

Now, we assume that the noise is time-independent, i.e. $\lambda=\text{const}$ and we have small $p_\text{data}\ll1$ such that $p_\text{data}\approx p\,(1+\lambda\dt)$. 
Then, minimizing the logical-error rate per unit physical time is equivalent to minimizing
\begin{equation}
    R(\lambda,\dt)=C\frac{(1+\lambda\dt)^{\alpha}}{\dt},
    \label{eq:static-rate-general}
\end{equation}
where we have constant 
\begin{equation}
    C=\frac{A}{d^\beta}\left(\frac{p}{\pth}\right)^{(d+1)/2}.
\end{equation}
We now compute the $\dt$ that gives minimal $R$. Taking the derivative of its logarithm gives
\begin{equation}
    \frac{\text{d}}{\text{d}\dt}\log R(\lambda,\dt)
    =\alpha\frac{\lambda}{1+\lambda\dt}-\frac1\dt .
\end{equation}
For $\alpha>1$, the unique unconstrained optimum is
\begin{equation}
    \dt^\star(\lambda)=\frac{1}{\lambda(\alpha-1)}\sim \frac{2}{\lambda gd},
    \label{eq:constant-opt}
\end{equation}
with minimal logical error rate per unit time
\begin{equation}
R^\star(\lambda)=C K_\alpha\lambda \sim e C\alpha\lambda
\end{equation}
where 
\begin{equation}
    K_\alpha=\frac{\alpha^\alpha}{(\alpha-1)^{\alpha-1}}\sim e\alpha
\end{equation}
where the right-hand sides are the limit of large $\alpha$.
Thus, the optimal syndrome measurement rate scales linearly with $\lambda$ and distance $d$. 
The improvement in logical error rate over fixed choice $\dt$ is given without approximations by
\begin{equation}\label{eq:gammaopt}
    \Gamma_{\rm opt}
    =\frac{R(\dt)}{R(\dt^\star)}=
    \frac{(1+\lambda\dt)^\alpha}{\lambda\dt K_\alpha}.
\end{equation}

Let us explicitly compare the logical error rates in the limit of large $d$.
The optimal choice $\dt^\star$ gives
\begin{equation}\label{eq:Roptimal}
R^\star(\lambda)\sim e\lambda g\frac{1}{2}\frac{A}{d^{\beta-1}}\left(\frac{p}{\pth}\right)^{d/2} 
\end{equation}
while with a constant syndrome interval $\dt$, we get
\begin{equation}
R(\lambda,\dt) \sim \frac{1}{\dt}\frac{ A }{d^\beta}\left(\frac{p}{\pth}\right)^{d/2} \left(1+\lambda\dt\right)^{gd/2}\,.
\end{equation}
We can directly read off the improvement in logical error rate as
\begin{equation}
    \Gamma_{\rm opt}=\frac{R(\dt)}{R(\dt^\star)}\sim\frac{2e^{-1}}{\lambda\dt gd}(1+\lambda\dt)^{gd/2}.
\end{equation}
Thus, we find an exponential improvement with $d$. 

We can also get a different view by considering $d\rightarrow \infty$ while keeping the ratio $\dt/\dt^\star$ fixed. 
To consider this limit, we  explicitly set $\lambda=1/(\dt^\star(\alpha-1))$, after which we get with some simple calculations
\begin{equation}
\Gamma_{\rm opt}=\frac{\alpha-1}{K_\alpha}\frac{\dt^\star}{\dt}\left(1+\frac{1}{\alpha-1}\frac{\dt}{\dt^\star}\right)^{\alpha}\sim \frac{1}{e}\frac{\dt^\star}{\dt}e^{\frac{\dt}{\dt^\star}}.
\end{equation}
Thus, we find that optimal $\dt^\star$ gives us an exponential improvement with increasing ratio $\dt/\dt^\star>1$, while for $\dt/\dt^\star<1$ a linear improvement in ratio $\dt^\star/\dt$.

\section{Derivation of adaptive advantage}\label{sec:adaptive_advantage}
We now show that adaptive protocols for syndrome intervals for bursts in $\lambda(t)$ have an advantage over all possible fixed-interval protocols.
We assume a burst protocol, where over a fraction $f$ of the total time we have idling noise rate $\lambda(t)=\lamh$, while all other times we have $\lambda(t)=\laml$.
We also define the burst noise ratio 
\begin{equation}
    r=\frac{\lamh}{\laml}.
\end{equation}

When the syndrome measurements are performed at fixed time-independent intervals $\dt$, we can optimize over all $\dt$ to the minimal logical error rate
\begin{equation}
    R_{\rm fixed}^\star=
    \min_{\dt}\left[(1-f)R(\laml,\dt)+f R(\lamh,\dt)\right]
    \label{eq:fixed-rate}
\end{equation}
where we use the logical error rate per unit of time according to our ansatz~\eqref{eq:ansatz_sup}.
In contrast, if we dynamically control $\dt(t)$ in time, then we can react to temporal changes in noise rate $\lambda(t)$ which yields lower logical error rates.
For the optimal time-dependent protocol, which applies the optimal interval $\dt^\star(\lambda(t))$ in time, the minimal logical error rate is given by
\begin{align}
    R_{\rm adapt}^\star&=
    \left[(1-f)\min_{\dt}R(\laml,\dt)+f\min_{\dt}R(\lamh,\dt)\right].
    \label{eq:oracle-rate-general}
\end{align}

First, let us compute the optimal adaptive logical error.
This is achieved by inserting the optimal measurement interval $\dt^\star(\lambda(t))$ of~\eqref{eq:oracle-rate-general} at every given time $t$, with which we get
\begin{equation}
    R_{\rm adapt}^\star\sim eC\alpha\laml[(1-f)+fr]\,.
\end{equation}

Next, let us compute $R_{\rm fixed}^\star$.
We write $x=\laml\dt$ which gives us
\begin{equation}
    R_{\rm fixed}^\star
    =C\laml\min_{x>0}
    \frac{(1-f)(1+x)^\alpha+f(1+rx)^\alpha}{x}.
    \label{eq:fixed-q}
\end{equation}
In the limit of large distance $\alpha\to\infty$  and setting $x=z/\alpha$ we have that
\begin{equation}
    (1+x)^\alpha\to e^z,
    \qquad
    (1+rx)^\alpha\to e^{rz}.
\end{equation}
We apply this to Eq.~\eqref{eq:fixed-q} and get
\begin{equation}
    R_{\rm fixed}^\star
    \sim C\laml\alpha\min_{z>0}
    \frac{(1-f)e^z+fe^{rz}}{z}.
    \label{eq:large-alpha-min}
\end{equation}

\subsection{Limit of small burst noise ratios}

Now, let us consider the limit of small burst noise ratios $r-1$. 
Let \(r=1+\epsilon\) with \(\epsilon\ll1\), and let
\(L\in\{1,r\}\) denote the instantaneous noise multiplier, with
\(L=1\) for a fraction \(1-f\) of the time and \(L=r\) for a
fraction \(f\).  Then
\[
\mu\equiv \mathbb E[L]=(1-f)+fr,
\qquad
\frac{{\rm Var}(L)}{\mu^2}
=
\frac{f(1-f)(r-1)^2}{[(1-f)+fr]^2}.
\]
The fixed-interval optimization can be written as
\[
R_{\rm fixed}^\star
\simeq
C\lambda_{\rm low}\alpha
\min_{z>0}
\frac{\mathbb E[e^{Lz}]}{z}.
\]
Setting \(L=\mu(1+\eta)\), with \(\mathbb E[\eta]=0\) and \(\mathbb E[\eta^2]=\text{Var}(L)/\mu^2\), and
\(z=y/\mu\), gives
\[
\frac{\mathbb E[e^{Lz}]}{z}
=
\mu\frac{e^y}{y}
\left[1+\frac{y^2}{2}\mathbb E[\eta^2]
+O((r-1)^3)\right].
\]
The minimum remains at \(y=1\) up to corrections that only affect the
value at higher order, so
\[
R_{\rm fixed}^\star
=
eC\lambda_{\rm low}\alpha\,\mu
\left[
1+\frac12
\frac{f(1-f)(r-1)^2}{[(1-f)+fr]^2}
+O((r-1)^3)
\right].
\]
By contrast,
\[
R_{\rm adapt}^\star
=
eC\lambda_{\rm low}\alpha\,\mu .
\]
Finally, we have
\[
\Gamma_{\rm adapt}
=
\frac{R_{\rm fixed}^\star}{R_{\rm adapt}^\star}
=
1+\frac12
\frac{f(1-f)(r-1)^2}{[(1-f)+fr]^2}
+O((r-1)^3).
\]
Thus, for small $r-1$, the advantage is controlled by the relative variance of the noise rate.
In the limit of $f\ll1$, we get
\[
\Gamma_{\rm adapt}
=
1+\frac12 f(r-1)^2
+O\!\left(f^2(r-1)^2,(r-1)^3\right).
\]

\subsection{Optimal burst noise advantage}

Next, we leave the small $r$ limit and find the optimal $r$ over all burst duration fractions $f$. For this, we optimize Eq.~\eqref{eq:large-alpha-min} over the burst duration fraction $f$, i.e. searching for the burst duration which gives the highest advantage. 
For Eq.~\eqref{eq:large-alpha-min}, the $z=z_\star$ that minimizes the equation solves
\begin{equation}
    (1-f)e^{z_\star}(z_\star-1)
    +f e^{rz_\star}(rz_\star-1)=0,
    \label{eq:zstar-eq}
\end{equation}
with $z_\star\in[1/r,1]$.  
This can be rewritten into
\begin{equation}
    \frac{(1-f)e^{z_\star}+fe^{rz_\star}}{z_\star}=
    (1-f)e^{z_\star}+fre^{rz_\star}.
    \label{eq:zstar-eq2}
\end{equation}
Combining both fixed and adaptive intervals and inserting stationary condition~\eqref{eq:zstar-eq2}, we get the advantage
\begin{equation}
    \Gamma_\text{adapt}(f,r)
    =\frac{R_{\rm fixed}^\star(f,r)}{R_{\rm adapt}^\star(f,r)}\sim\frac{(1-f)e^{z_\star}+fr e^{rz_\star}}
          {e[(1-f)+fr]}.
    \label{eq:gamma-large-alpha}
\end{equation}
For \(1/r<z^\star<1\), we can rewrite the stationary condition~\eqref{eq:zstar-eq}
\begin{equation}
    \frac{f}{1-f}=e^{-(r-1)z}\frac{1-z}{rz-1}.
\end{equation}
This condition on $f$ is then inserted into~\eqref{eq:gamma-large-alpha} to give
\begin{equation}
    \Gamma(z,r)=\frac{(r-1)e^{z-1}}{rz-1+re^{-(r-1)z}(1-z)}
\end{equation}
and by taking the logarithm and performing the remaining one-variable optimization over $r$  yields the optimal saddle point as
\begin{equation}
    z_{\rm opt}=\frac{\log r}{r-1},
    \label{eq:zopt}
\end{equation}
with
\begin{equation}
    f^\star(r)=
    \frac{r-1-\log r}
         {(r^2-1)\log r-(r-1)^2},
    \label{eq:fstar_sup}
\end{equation}
Finally,
\begin{equation}
    \Gamma_\text{adapt}(f^\star,r)
    =\frac{r-1}{e\log r}\,r^{1/(r-1)}
    .
    \label{eq:gamma-max_sup}
\end{equation}
For large burst ratios $r$ we get
\begin{equation}
    \Gamma_\text{adapt}(f^\star,r)\sim \frac{r}{e\log r},
    \qquad
    f^\star(r)\sim \frac{1}{r\log r}.
    \label{eq:large-r_sup}
\end{equation}
Thus, short but strong bursts provide the largest benefit with a sublinear scaling with $r$.  For example, we get $\Gamma_\text{adapt}(10)\simeq1.86$ and $\Gamma_\text{adapt}(20)\simeq2.73$.

\section{Detector-likelihood burst controller}
\label{sec:finite_window_detection}
For time-dependent $\lambda(t)$, we  adaptively control $\dt$ depending on the syndrome activity. As the syndrome activity itself is noisy due to shot noise, we need a criterion to distinguish the higher burst-noise from random fluctuations.
We use a likelihood-ratio test between the two hypotheses
\begin{equation}
    H_0:\lambda=\lambda_{\rm low},
    \qquad
    H_1:\lambda=\lambda_{\rm high}.
\end{equation}

In the phenomenological model, at round $j$ we measure the outcome of each stabilizer check $c$ with outcome $s_{j,c}$.
The detector bit quantifies the change in syndrome every round via
\begin{equation}
    D_{j,c}=s_{j,c}\oplus s_{j-1,c}.
\end{equation}
In the phenomenological model, the detector clicks when an odd number of relevant data and measurement faults touch that detector.  If the relevant fault probabilities are collected in a set $\{p_e\}_{e\in\partial(j,c)}$, then the probability of observing a change in detector is given by
\begin{equation}
    q_c(\lambda,\dt_j)
    =P(D_{j,c}=1\mid \lambda,\dt_j)
    =\frac12\left[1-\prod_{e\in\partial(j,c)}(1-2p_e)\right].
    \label{eq:S-click-prob}
\end{equation}
This expression automatically accounts for bulk and boundary checks, because boundary checks have fewer incident data edges.

The per-round log-likelihood ratio is
\begin{equation}
\ell_j=\frac{1}{N_\text{c}}\sum_c\left[
D_{j,c}\log\frac{q_c(\lambda_{\rm high},\dt_j)}{q_c(\lambda_{\rm low},\dt_j)}
+(1-D_{j,c})\log\frac{1-q_c(\lambda_{\rm high},\dt_j)}{1-q_c(\lambda_{\rm low},\dt_j)}
\right].
\label{eq:S-llr}
\end{equation}
The normalization by $N_\text{c}$ makes thresholds comparable across distances and sectors.  

To further reduce shot-noise, we average over a window of $W$ syndrome measurement rounds
\begin{equation}
    \bar\ell^{(W)}=\frac1W\sum_{i=0}^{W-1}\ell_{j-i}.
\end{equation}
It does not trigger during the first $W-1$ rounds. Once $W$ rounds are available, it switches to the burst interval if
\begin{equation}
    \bar\ell^{(W)}\ge \theta .
\end{equation}
After a trigger, the smaller burst interval $\dt_\text{high}$ is used for a fixed time $fT$, where $f$ is the burst duration fraction. After that, we switch back to $\dt_\text{low}$.

For calibration, one may approximate the low- and high-state
distributions by Gaussian distributions
\begin{equation}
    \bar\ell^{(W)}\mid H_a
    \approx
    \mathcal{N}\left(
        \mu_a,\frac{\sigma_{a,\rm eff}^2}{W}
    \right),
    \qquad a\in\{0,1\},
\end{equation}
where $\sigma_{a,\rm eff}^2$ includes the finite-range temporal
correlations between neighboring rounds.

Then we have the false positives and false negatives
\begin{align}
    P_{\rm FP}(W,\theta)&=P(\bar\ell^{(W)}\ge\theta\mid H_0),\\
    P_{\rm FN}(W,\theta)&=P(\bar\ell^{(W)}<\theta\mid H_1),
\end{align}
respectively.

The window length $W$ should not be made arbitrarily large: increasing $W$ improves statistical confidence but also adds an inference latency of roughly $W\dt_{\rm n}$.  Thus $W$ and $\theta$ are chosen to balance false positives, missed bursts, and inference latency.  In the numerical scans we optimize these parameters directly against the logical failure probabilities.%

False positives waste rounds in the burst interval during quiet periods, false negatives leave the memory exposed to high data noise, and a finite detection window produces inference latency.

\subsection{Explicit bound}

We now give a finite-sample bound on the reliability of the burst
detector.  %
Changes in the physical error rate can be inferred directly
from syndrome statistics~\cite{tan2026resilience}; in particular,
Ref.~\cite{tan2026resilience} derived exponentially small
misclassification probabilities for a likelihood-based detector
constructed from an independent subset of syndrome variables.
Here we establish an analogous exponential concentration result for
the detector-wise score used by our controller, while retaining the
full set of locally correlated detectors.

We consider a window of $W$ rounds that is
entirely contained in one of the two noise states $H_0$ or $H_1$, and
assume that the probing interval remains $\Delta t_{\rm low}$ throughout
this window.  A window that straddles a transition between the two noise
states instead contributes to the detection latency.

We define
\begin{equation}
q_{a,c}
=
q_c(\lambda_a,\Delta t_{\rm low}),
\qquad
a\in\{0,1\},
\end{equation}
where
\(\lambda_0=\lambda_{\rm low}\) and
\(\lambda_1=\lambda_{\rm high}\).
For detector $c$, define the detector-wise log-likelihood contribution
\begin{equation}
X_c(D)
=
D\log\frac{q_{1,c}}{q_{0,c}}
+
(1-D)\log\frac{1-q_{1,c}}{1-q_{0,c}} .
\label{eq:single_detector_llr}
\end{equation}
The statistic used by the controller can then be written as
\begin{equation}
\bar\ell_j^{(W)}
=
\frac{1}{W N_\text{c}}
\sum_{k=j-W+1}^{j}\sum_{c=1}^{N_\text{c}}
X_c(D_{k,c}).
\label{eq:window_llr_exp}
\end{equation}
Because neighboring detector variables are correlated,
Eq.~\eqref{eq:window_llr_exp} is in general a composite
log-likelihood score rather than the exact joint log-likelihood ratio.
The correlations will be explicitly accounted for below.

The mean score under the high-noise hypothesis is positive,
whereas that under the low-noise hypothesis is negative:
\begin{align}
\mu_1
&=
\mathbb{E}\!\left[\bar\ell_j^{(W)}\middle|H_1\right]
=
\frac{1}{N_\text{c}}\sum_{c=1}^{N_\text{c}}
D_{\rm KL}\!\left(
{\rm Ber}(q_{1,c})
\,\Vert\,
{\rm Ber}(q_{0,c})
\right),
\\
\mu_0
&=
\mathbb{E}\!\left[\bar\ell_j^{(W)}\middle|H_0\right]
=
-\frac{1}{N_\text{c}}\sum_{c=1}^{N_\text{c}}
D_{\rm KL}\!\left(
{\rm Ber}(q_{0,c})
\,\Vert\,
{\rm Ber}(q_{1,c})
\right),
\end{align}
where $\text{Ber}(q)$ is a Bernoulli random variable with probability $q$ and 
\begin{equation}
D_{\rm KL}\!\left(
{\rm Ber}(q)\,\Vert\,{\rm Ber}(q')
\right)
=
q\log\frac{q}{q'}
+
(1-q)\log\frac{1-q}{1-q'}.
\end{equation}
Consequently, their separation is
\begin{align}
\Delta\mu
&\equiv
\mu_1-\mu_0
\nonumber\\
&=
\frac{1}{N_\text{c}}\sum_{c=1}^{N_\text{c}}
\Big[
D_{\rm KL}\!\left(
{\rm Ber}(q_{1,c})
\,\Vert\,
{\rm Ber}(q_{0,c})
\right)
+
D_{\rm KL}\!\left(
{\rm Ber}(q_{0,c})
\,\Vert\,
{\rm Ber}(q_{1,c})
\right)
\Big].
\label{eq:llr_gap}
\end{align}
Thus $\Delta\mu>0$ whenever the two noise states produce
distinguishable detector statistics.

We next establish exponential concentration of
$\bar\ell_j^{(W)}$.
For each detector define the two possible values
\begin{equation}
x_{1,c}
=
\log\frac{q_{1,c}}{q_{0,c}},
\qquad
x_{0,c}
=
\log\frac{1-q_{1,c}}{1-q_{0,c}},
\end{equation}
and their range
\begin{equation}
B_c
=
|x_{1,c}-x_{0,c}|
=
\left|
\log
\frac{q_{1,c}(1-q_{0,c})}
     {q_{0,c}(1-q_{1,c})}
\right|.
\label{eq:Bc}
\end{equation}
We also define
\begin{equation}
\overline{B^2}
=
\frac{1}{N_\text{c}}\sum_{c=1}^{N_\text{c}}B_c^2.
\label{eq:Bbar}
\end{equation}

The crucial observation is that the detector variables have only
local dependence.  In the phenomenological model, let $E_{j,c}$
denote the parity of data faults incident on check $c$ between
rounds $j-1$ and $j$, and let $M_{j,c}$ denote the measurement
fault of that check in round $j$.  The detector can be written as
\begin{equation}
D_{j,c}
=
E_{j,c}\oplus M_{j,c}\oplus M_{j-1,c}.
\label{eq:local_detector_support}
\end{equation}
Hence every $D_{j,c}$, and therefore every
$X_c(D_{j,c})$, depends only on a bounded set of local primitive
fault variables.

Consider the graph $G_W$ whose vertices are the $W N_\text{c}$ detector
variables in the window, with an edge between two vertices whenever
their underlying sets of primitive fault variables overlap.
Because the primitive faults are independent conditional on $H_a$,
$G_W$ is a dependency graph: any two sets of vertices with no edge
between them depend on disjoint sets of primitive faults and are
therefore independent.  Locality implies that its maximum degree
$\nu$ is bounded independently of $d$ and $W$.
For the phenomenological surface-code detector graph used here,
a bulk detector has only a constant number of spatial and temporal
neighbors; the precise value of this constant will not be important.

We now prove a Hoeffding-type concentration inequality for this
dependency graph.  Let
\begin{equation}
S
=
\sum_{k=j-W+1}^{j}\sum_{c=1}^{N_\text{c}}
X_c(D_{k,c}),
\qquad
N=W N_\text{c} ,
\end{equation}
so that $\bar\ell_j^{(W)}=S/N$.
A graph of maximum degree $\nu$ can be properly colored with
\begin{equation}
\chi\le \nu+1
\end{equation}
colors.  Let $\mathcal{C}_1,\ldots,\mathcal{C}_\chi$ denote the
corresponding color classes.  Variables belonging to the same
color class are mutually independent.

Writing
\begin{equation}
Y_v=X_v-\mathbb{E}X_v
\end{equation}
for a detector contribution indexed by vertex $v$, Hoeffding's
lemma gives
\begin{equation}
\mathbb{E}\!\left[e^{uY_v}\right]
\le
\exp\left(\frac{u^2 B_v^2}{8}\right).
\label{eq:hoeffding_lemma_detector}
\end{equation}
Using H\"older's inequality across color classes and independence
within each color class, for any $u>0$ we obtain
\begin{align}
\mathbb{E}\!
\left[
e^{u(S-\mathbb{E}S)}
\right]
&=
\mathbb{E}\!
\left[
\prod_{r=1}^{\chi}
e^{u\sum_{v\in\mathcal{C}_r}Y_v}
\right]
\nonumber\\
&\le
\prod_{r=1}^{\chi}
\left[
\mathbb{E}
e^{\chi u\sum_{v\in\mathcal{C}_r}Y_v}
\right]^{1/\chi}
\nonumber\\
&=
\prod_{r=1}^{\chi}
\prod_{v\in\mathcal{C}_r}
\left[
\mathbb{E}e^{\chi uY_v}
\right]^{1/\chi}
\nonumber\\
&\le
\exp\left[
\frac{\chi u^2}{8}
\sum_v B_v^2
\right].
\label{eq:dependency_mgf}
\end{align}
A Chernoff bound therefore gives, for any $s>0$,
\begin{align}
{\rm Pr}(S-\mathbb{E}S\ge s)
&\le
\inf_{u>0}
\exp\left[
-us+
\frac{\chi u^2}{8}\sum_v B_v^2
\right]
\nonumber\\
&=
\exp\left[
-\frac{2s^2}
{\chi\sum_v B_v^2}
\right].
\label{eq:dependency_hoeffding}
\end{align}
The same argument applied to $-S$ gives the corresponding lower-tail
bound.

Since each spatial detector type appears once in each of the $W$
rounds,
\begin{equation}
\sum_v B_v^2
=
W\sum_{c=1}^{N_\text{c}}B_c^2
=
W N_\text{c}\,\overline{B^2}.
\end{equation}
Using $\chi\le\nu+1$ and
$s=W N_\text{c}\delta$, Eq.~\eqref{eq:dependency_hoeffding} yields
\begin{equation}
{\rm Pr}\!\left(
\bar\ell_j^{(W)}-\mu_a\ge\delta
\middle|H_a
\right)
\le
\exp\left[
-\frac{2W N_\text{c}\delta^2}
{(\nu+1)\overline{B^2}}
\right],
\label{eq:upper_exp_concentration}
\end{equation}
with the analogous bound for the lower tail.

It follows that, for any threshold
$\mu_0<\theta<\mu_1$,
\begin{align}
P_{\rm FP}(W,\theta)
&\le
\exp\left[
-\frac{2W N_\text{c}(\theta-\mu_0)^2}
{(\nu+1)\overline{B^2}}
\right],
\label{eq:pfp_exp}
\\
P_{\rm FN}(W,\theta)
&\le
\exp\left[
-\frac{2W N_\text{c}(\mu_1-\theta)^2}
{(\nu+1)\overline{B^2}}
\right].
\label{eq:pfn_exp}
\end{align}

For the symmetric threshold
\begin{equation}
\theta_{\rm mid}
=
\frac{\mu_0+\mu_1}{2},
\end{equation}
both error probabilities obey
\begin{equation}
\max\{P_{\rm FP},P_{\rm FN}\}
\le
\exp\left[
-\frac{W N_\text{c}(\Delta\mu)^2}
{2(\nu+1)\overline{B^2}}
\right].
\label{eq:midpoint_exp_bound}
\end{equation}

A simpler, although more conservative, expression follows by assuming
that the detector click probabilities are bounded away from zero and
one,
\begin{equation}
\eta\le q_{a,c}\le1-\eta
\qquad
\text{for all }a,c,
\label{eq:eta_bound}
\end{equation}
where $\eta>0$ is independent of $d$ and $W$.  Defining
\begin{equation}
L_\eta=\log\frac{1-\eta}{\eta},
\end{equation}
we have
\begin{equation}
|x_{0,c}|,|x_{1,c}|\le L_\eta,
\qquad
B_c\le2L_\eta,
\qquad
\overline{B^2}\le4L_\eta^2.
\end{equation}
For one CSS sector of the odd-distance rotated surface code,
\begin{equation}
N_\text{c}=\frac{d^2-1}{2}.
\end{equation}
Equation~\eqref{eq:midpoint_exp_bound} therefore implies
\begin{equation}
\max\{P_{\rm FP},P_{\rm FN}\}
\le
\exp\left[
-\frac{
W(d^2-1)(\Delta\mu)^2
}{
16(\nu+1)L_\eta^2
}
\right].
\label{eq:exponential_detector_bound}
\end{equation}
Thus, provided that the two detector distributions remain separated
by a $d$-independent $\Delta\mu>0$, and $\eta$ and $\nu$ remain
$d$-independent, the misclassification probabilities satisfy
\begin{equation}
P_{\rm FP},P_{\rm FN}
=
\exp[-\Omega(Wd^2)].
\label{eq:exp_scaling_detector}
\end{equation}

In particular, to make both error probabilities at most $\epsilon$,
it is sufficient to choose
\begin{equation}
W
\ge
\frac{
16(\nu+1)L_\eta^2
}{
(d^2-1)(\Delta\mu)^2
}
\log\frac{1}{\epsilon}.
\label{eq:window_length_exp}
\end{equation}
Hence the number of rounds required for a prescribed confidence
decreases as
$O(\log(1/\epsilon)/d^2)$ until the obvious lower limit $W=1$ is
reached.

\section{Numerical verification of optimal syndrome interval}\label{sec:fixedt}
\begin{figure}[htpb]
\centering
\subfigimg[width=0.3\columnwidth]{}{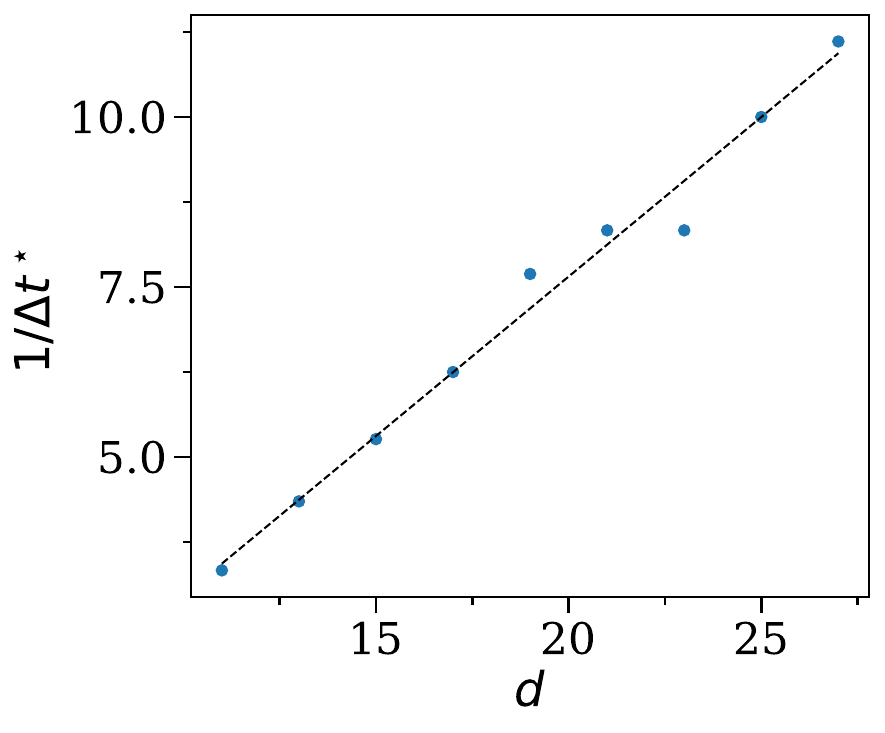}
\caption{Optimal syndrome interval $\dt^\star$ for quantum error correction memory for time-independent idle noise against distance $d$. We have $\lambda=1$, $p=0.015$ and $T=200$. $\dt^\star$ is found from  simulation of phenomenological noise model for different code distances $d$ of rotated surface codes. Dashed line is fit with linear ansatz $1/\dt^\star(d)= \eta d+z$ with $\eta=0.47$. 
}
\label{fig:fixedt}
\end{figure}

We numerically determine the optimal syndrome timing for constant noise in Fig.~\ref{fig:fixedt}. We find the interval $\dt^\star$ with the minimal $\pL$ for the full numerical simulation of the phenomenological model for different distances $d$ as shown in the main text. As simulation parameters, we have time-independent idle rate $\lambda=1$, base noise scale $p=0.015$, readout error factor $b_\text{read}=1$ and total time $T=200$. 
We then fit with linear ansatz $1/\dt^\star(d)= \eta d+z$, demonstrating the inverse relationship between $dt^\star$ and $d$. 

\section{Mapping to Google Willow experiment}
\label{sec:supp_google_mapping}

In this section we estimate the parameters of our phenomenological
idle-noise model using the recent Google Willow surface-code
experiment~\cite{google2025quantum}. Our goal is not to reproduce the
full circuit-level noise model, but to obtain an order-of-magnitude
mapping between the experimentally reported component error budget and
the parameters of our analytic theory.

The Google suppression factor \(\Lambda\) is extracted from the empirical
distance scaling
\[
    p_L(d)\propto \Lambda^{-(d+1)/2}.
\]
Equivalently, \(\Lambda^{-1}\) is the effective error-suppression
parameter per half-distance increment. In the component error budget, this
quantity is decomposed approximately into sensitivity-weighted error
contributions. If \(p_i\) is the microscopic probability of error source
\(i\), and \(w_i\) is the fitted sensitivity of the logical error to that
source, we write
\[
    C_i=w_i p_i.
\]
Thus \(C_i\) is not the bare physical error probability, but the
contribution of that component to the effective inverse suppression
factor, normalized by the threshold scale. To leading order one may view
the experimentally observed inverse suppression factor as
\[
    \Lambda^{-1}
    \simeq
    \sum_i C_i ,
\]
up to residual circuit-level effects, correlations, leakage dynamics,
non-ideal decoder, and finite-distance corrections. For the Google
Willow experiment, the reported value is
\[
    \Lambda_{\rm Google}\simeq2.14,
\]
which is close to the reported $\sum_iC_i$.
Thus, we use the calibration $C_i$ to assign relative phenomenological noise
parameters. %

\begin{table}[h]
\centering
\begin{tabular}{l|c|c|c}
\hline
Component & Symbol & \(C_i\) & Assignment \\
\hline
CZ gate errors 
& \(C_{\rm CZ}\) 
& 0.182 
& \(C_{\rm gate}^{\rm data}\) \\
CZ crosstalk 
& \(C_{\rm CZ,xtalk}\) 
& 0.050 
& \(C_{\rm gate}^{\rm data}\) \\
CZ leakage 
& \(C_{\rm CZ,leak}\) 
& 0.022 
& \(C_{\rm gate}^{\rm data}\) \\
single-qubit gates 
& \(C_{\rm SQ}\) 
& 0.039 
& \(C_{\rm gate}^{\rm data}\) \\
\hline
data idle 
& \(C_{\rm idle}^{\rm data}\) 
& 0.090 
& \(\lambda p\dt\) \\
\hline
\hline
total listed contribution 
& \(\sum_i C_i\) 
& 0.440 
& \(\Lambda^{-1}\) estimate \\
\hline
\end{tabular}
\caption{
Sensitivity-weighted Google error-budget components and their assignment
to the phenomenological parameters used in our model. The \(C_i\) are
dimensionless contributions to the effective inverse suppression factor,
\(C_i=w_i p_i/p_{\rm th}\), rather than bare microscopic error
probabilities.
}
\label{tab:google_error_budget_assignment}
\end{table}

In the
Google experiment, one complete syndrome-extraction cycle has duration
\[
    \tau_{\rm cyc}\simeq 1.108\,\mu{\rm s}.
\]
We choose units such that
\[
    \dt=1
\]
corresponds to this full Google syndrome-cycle time. Thus \(\dt\)
denotes the physical time between consecutive syndrome rounds, measured
in units of \(\tau_{\rm cyc}\). With this convention, the runtime of the
phenomenological model can be compared directly with the experimental
cycle time.

The Google error budget contains a data-qubit idle contribution. This
term should be interpreted as the sensitivity-weighted contribution of
idle locations that occur inside the scheduled syndrome-extraction
circuit, rather than as an independently inserted full-cycle storage
interval. We therefore use the reported data-idle budget term directly to
calibrate the coefficient of the time-dependent idle contribution in our
phenomenological model.

We decompose the data-like contribution into a non-idle part and an idle
part. The non-idle data-like contribution contains two-qubit gate errors,
crosstalk, leakage, and single-qubit gate errors,
\begin{align}
    C_{\rm gate}^{\rm data}
    &=
    C_{\rm CZ}
    +C_{\rm CZ,xtalk}
    +C_{\rm CZ,leak}
    +C_{\rm SQ}
    \nonumber\\
    &=
    0.182+0.050+0.022+0.039
    \nonumber\\
    &=
    0.293 .
    \label{eq:Cgate_data_google}
\end{align}
As these are related to the gates for the syndrome measurement itself, their contribution is independent of waiting time.

In contrast, the reported data-qubit idle contribution is
\[
    C_{\rm idle}^{\rm data}=0.090 .
\]
We absorb the non-idle data-like contribution into the definition of the
phenomenological scale \(p\),
\[
    p \equiv C_{\rm gate}^{\rm data} p_{\rm th}.
\]
The idle contribution increases with $\dt$ and is then modeled as the \(\lambda\dt\) term,
\[
    p_{\rm data}(\dt)
    \simeq
    p\left(1+\lambda\dt\right),
    \label{eq:pdata_google_mapping}
\]
where \(\dt=1\) corresponds to one Google syndrome-cycle time.
Matching the reported data-idle contribution at \(\dt=1\) gives
\[
    \lambda_{\rm Google}
    =
    \frac{C_{\rm idle}^{\rm data}}
    {C_{\rm gate}^{\rm data}}
    =
    \frac{0.090}{0.293}
    \simeq
    0.31 .
    \label{eq:lambda_google}
\]
With this convention, the total data-like contribution at the actual
Google cycle time is
\[
    C_{\rm gate}^{\rm data}
    \left(1+\lambda_{\rm Google}\right)
    \simeq
    0.293(1+0.31)
    \simeq
    0.383,
\]
which reproduces the total data-like cost of one Google syndrome cycle,
\[
    C_{\rm data}^{\rm Google}
    =
    C_{\rm gate}^{\rm data}
    +
    C_{\rm idle}^{\rm data}
    =
    0.293+0.090
    =
    0.383 .
\]
Thus the Google-inspired phenomenological parameters of our model are
\[
    p/p_{\rm th}\simeq0.293,
    \qquad
    \lambda_{\rm Google}\simeq0.31.
\]

With this, we compute the explicit $\Gamma_\text{opt}$ improvement  in logical error rate via~\eqref{eq:gammaopt} with our estimated Google Willow parameters in Table~\ref{tab:google_timings_unconstrained}. We compute the optimal $\dt^\star$, improvement $\Gamma_\text{opt}$ and optimal physical wait time $\dt^\star$ in units of $\mu s$. For~\eqref{eq:gammaopt}, we assume $g=0.8$ as fitted from our simulations of the rotated surface code.

\begin{table}[t]
\centering
\begin{tabular}{c|cccccccccccc}
\hline
\(d\) & 3 & 5 & 7 & 9 & 11 & 13 & 15 & 17 & 19 & 21 & 23 & 27 \\
\hline
\hline
\(\dt^\star\)
& 5.38 & 2.30 & 1.47 & 1.08 & 0.85 & 0.70 & 0.60 & 0.52 & 0.46 & 0.41 & 0.38 & 0.32 \\
\hline
\(\dt^\star\) [\(\mu\)s]
& 5.96 & 2.55 & 1.62 & 1.19 & 0.94 & 0.78 & 0.66 & 0.58 & 0.51 & 0.46 & 0.42 & 0.35 \\
\hline
\(\Gamma_{\rm opt}\)
& 1.72 & 1.21 & 1.05 & 1.00 & 1.01 & 1.06 & 1.13 & 1.24 & 1.37 & 1.54 & 1.74 & 2.28 \\
\hline
\end{tabular}

\label{tab:google_timings_unconstrained}
\caption{
Estimated optimal syndrome intervals for the Google-inspired parameters
\(\lambda=0.31\) and \(g=0.8\). The interval \(\dt^\star\) is shown
in units of one Google syndrome-cycle time,
\(\tau_{\rm cyc}=1.108\,\mu{\rm s}\) and units of $\mu s$. The improvement factor is computed
relative to the experimental baseline \(\dt_0=1\).
}
\end{table}

We find that the experimental $\dt$ is already nearly optimal at $d=9$, while $\dt$ is too small for $d<9$, while the original $\dt$ would be too large for $d>9$.
The experiment has been performed for $d=3,5,7$, which implies that a larger $\dt$ for these $d$ would directly yield lower logical error rates. For example, for $d=3$, we find $\Gamma_\text{opt}=1.72$, which corresponds to a relative reduction of logical error rate $1-1/\Gamma_\text{opt}\approx 0.4$.
We note that as $\dt=1$ corresponds to the current time of the measurement cycle itself, implementing $\dt<1$ would require faster stabilizer measurement speeds than in the current experiment.

\section{Effective noise suppression via optimized intervals}\label{sec:dist_reduction}
We now consider the effective noise reduction achieved by our optimized syndrome intervals. 

We now quantify the advantage in the limit of large $d$ obtained from using the optimized
syndrome interval.  First, let us recall from Sec.~\ref{sec:optimal} that for a fixed interval \(\Delta t\), the logical error rate per unit of time is
\begin{equation}
    R_{\rm fixed}(d,\Delta t)
    =
    \frac{A}{d^\beta}
    \left(\frac{p}{p_{\rm th}}\right)^{(d+1)/2}
    \frac{(1+\lambda\Delta t)^{\alpha(d)}}{\Delta t},
    \qquad
    \alpha(d)=g\frac{d+1}{2}.
    \label{eq:R_fixed_distance}
\end{equation}
For constant \(\lambda\), the optimized interval is
\begin{equation}
    \Delta t^\star(d)
    =
    \frac{1}{\lambda(\alpha(d)-1)},
\end{equation}
and therefore
\begin{equation}
    R_{\rm opt}(d)
    =
    \frac{A}{d^\beta}
    \left(\frac{p}{p_{\rm th}}\right)^{(d+1)/2}
    \lambda\,K_{\alpha(d)},
    \qquad
    K_\alpha
    =
    \frac{\alpha^\alpha}{(\alpha-1)^{\alpha-1}} .
    \label{eq:R_opt_distance}
\end{equation}
One can describe the improvement in terms of an effective
distance-scan suppression factor.  Keeping only the leading exponential
dependence on \(d\), Eq.~\eqref{eq:R_fixed_distance} gives
\begin{equation}
    R_{\rm fixed}(d,\Delta t)
    =
    {\rm poly}(d)\,
    \left[
        (1+\lambda\Delta t)^g
        \frac{p}{p_{\rm th}}
    \right]^{(d+1)/2},
\end{equation}
where \({\rm poly}(d)\) denotes algebraic and nonuniversal prefactors.  By
contrast, Eq.~\eqref{eq:R_opt_distance} gives
\begin{equation}
    R_{\rm opt}(d)
    =
    {\rm poly}(d)\,
    \left(
        \frac{p}{p_{\rm th}}
    \right)^{(d+1)/2}.
\end{equation}
Therefore the leading exponential suppression factors are
\begin{equation}
    \Lambda_{\rm fixed}
    =
    \frac{p_{\rm th}}
    {(1+\lambda\Delta t)^g p},
    \qquad
    \Lambda_{\rm opt}
    =
    \frac{p_{\rm th}}{p}.
    \label{eq:lambda_fixed_opt}
\end{equation}
Thus optimized timing increases suppression factor by
\begin{equation}
    \frac{\Lambda_{\rm opt}}{\Lambda_{\rm fixed}}
    =
    (1+\lambda\Delta t)^g .
    \label{eq:lambda_improvement}
\end{equation}

For example, using the $\Lambda_{\rm Google}\simeq2.14$ reported in Ref.~\cite{google2025quantum} and the estimated $\lambda\simeq0.31$ from Sec.~\ref{sec:supp_google_mapping}, we get
\begin{equation}
    g\log(1+\lambda_{\rm Google}\Delta t)
    =
    0.8\log(1.31)
    \simeq
    0.216.
\label{eq:google_distance_saving_numerator}
\end{equation}
where we use our fitted exponent fraction $g\simeq 0.8$. Then, we get for the noise suppression factor %
\begin{equation}
    \Lambda^\text{opt}_{\rm Google}\simeq2.66,
\end{equation}
a $24\%$ increase over the fixed-interval $\Lambda_{\rm Google}$.
Note that we discarded polynomial factors in $R_\text{fixed}$ and $R_\text{opt}$. As $R_\text{opt}$ has an additional factor $d$ compared to $R_\text{fixed}$, for small $d$ the actual advantage may be lower. 
Additionally, the optimized $\dt^\star\propto 1/d$ requires the ability to implement faster syndrome measurements with increasing distance.

\subsection{Distance reduction via optimized intervals}
Next, we ask how much distance can one save by using the optimized interval while preserving the logical error, i.e. what is
\(d_{\rm opt}\) that gives the same logical error as a fixed-timing code of distance
\(d_{\rm fixed}\).  
This condition is
\begin{equation}
    R_{\rm opt}(d_{\rm opt})
    =
    R_{\rm fixed}(d_{\rm fixed},\Delta t).
    \label{eq:distance_saving_full_condition}
\end{equation}
Using Eqs.~\eqref{eq:R_fixed_distance} and~\eqref{eq:R_opt_distance}, this gives
the implicit finite-distance equation
\begin{equation}
    \lambda\,K_{\alpha(d_{\rm opt})}\,
    d_{\rm opt}^{-\beta}
    \Lambda_{\rm opt}^{-(d_{\rm opt}+1)/2}
    =
    \Delta t^{-1}\,
    d_{\rm fixed}^{-\beta}
    \Lambda_{\rm fixed}^{-(d_{\rm fixed}+1)/2}.
    \label{eq:dopt_full_implicit}
\end{equation}
Taking logarithms,
\begin{equation}
    d_{\rm opt}+1
    =
    \frac{
        (d_{\rm fixed}+1)\log\Lambda_{\rm fixed}
        +2\beta\log(d_{\rm fixed}/d_{\rm opt})
        +2\log[\lambda\Delta t\,K_{\alpha(d_{\rm opt})}]
    }
    {\log\Lambda_{\rm opt}} .
    \label{eq:dopt_full_log}
\end{equation}
The last two terms are only logarithmic in distance.  Hence, in the asymptotic
distance-scaling limit, the algebraic prefactors drop out and one obtains
\begin{equation}
    d_{\rm opt}+1
    =
    (d_{\rm fixed}+1)
    \frac{\log\Lambda_{\rm fixed}}
         {\log\Lambda_{\rm opt}}
    \label{eq:dopt_asymptotic}
\end{equation}
where we suppressed terms of order $O(\log d_{\rm fixed})$.
Equivalently,
\begin{equation}
    d_{\rm opt}
    =
    (d_{\rm fixed}+1)
    \frac{\log\Lambda_{\rm fixed}}
         {\log\Lambda_{\rm opt}}
    -1.
    \label{eq:dopt_general}
\end{equation}
The asymptotic distance saving is therefore
\begin{equation}
    \Delta d
    \equiv
    d_{\rm fixed}-d_{\rm opt}
    =
    (d_{\rm fixed}+1)
    \left[
        1-
        \frac{\log\Lambda_{\rm fixed}}
             {\log\Lambda_{\rm opt}}
    \right].
    \label{eq:delta_d_general}
\end{equation}
Using
\(\Lambda_{\rm opt}=\Lambda_{\rm fixed}(1+\lambda\Delta t)^g\), this can also
be written as
\begin{equation}
    \Delta d
    =
    (d_{\rm fixed}+1)
    \left[
        1-
        \frac{\log\Lambda_{\rm fixed}}
        {\log\Lambda_{\rm fixed}
        +g\log(1+\lambda\Delta t)}
    \right].
    \label{eq:delta_d_fixed_form}
\end{equation}

\section{Coherence-time estimate of the idle contribution}
\label{sec:supp_google_pmem_check}

We estimate the
memory error accumulated over one syndrome-cycle time directly from standard noise metrics, and apply this for the Google Willow QEC device. 

We estimate the idle-memory error by Pauli-twirling a single-qubit
amplitude-relaxation and dephasing channel. Let the resulting Pauli channel
have probabilities \(p_X,p_Y,p_Z\), where we simply count whether a fault occurred. The corresponding Bloch-axis
eigenvalues are
\[
    \lambda_X=1-2(p_Y+p_Z),\qquad
    \lambda_Y=1-2(p_X+p_Z),\qquad
    \lambda_Z=1-2(p_X+p_Y).
\]
For a \(T_1/T_2\) memory channel over a time \(\tau\),
\[
    \lambda_X=\lambda_Y=e^{-\tau/T_2},
    \qquad
    \lambda_Z=e^{-\tau/T_1}.
\]
Assuming \(p_X=p_Y=q\) and \(p_Z=r\), we obtain
\[
    q=\frac{1-e^{-\tau/T_1}}{4},
    \qquad
    r=
    \frac{1-e^{-\tau/T_2}}{2}
    -
    \frac{1-e^{-\tau/T_1}}{4}.
\]
The total non-identity Pauli probability is therefore
\[
    p_{\rm mem}(\tau)
    =p_X+p_Y+p_Z 
    =2q+r 
    =
    \frac{3}{4}
    -
    \frac{1}{4}e^{-\tau/T_1}
    -
    \frac{1}{2}e^{-\tau/T_2}.
\]
For short times, we have 
\[
\begin{aligned}
p_{\rm mem}(\tau)\sim
    \frac{\tau}{4T_1}
    +
    \frac{\tau}{2T_2}.
\end{aligned}
\]

The Willow QEC device has a reported
mean relaxation time
\[
    T_1 \simeq 68\,\mu{\rm s},
\]
dephasing time
\[
    T_2\simeq 89\,\mu{\rm s},
\]
and a surface-code cycle rate of approximately \(909{,}000\) cycles per
second, corresponding to
\[
    \tau_{\rm cyc}\simeq 1.108\,\mu{\rm s}.
\]
The \(T_1\), \(T_2\), cycle-rate, and \(\Lambda_{3,5,7}=2.14\pm0.02\) values are
reported in Ref.~\cite{google2025quantum}.

For one Google syndrome-cycle time,
\[
    \tau=\tau_{\rm cyc}\simeq 1.108\,\mu{\rm s},
\]
we find 
\[
\begin{aligned}
    p_{\rm mem}(\tau_{\rm cyc})
    &=
    \frac{3}{4}
    -
    \frac{1}{4}e^{-1.108/68}
    -
    \frac{1}{2}e^{-1.108/89}
    \\
    &\simeq
    1.02\times 10^{-2}.
\end{aligned}
\]
Using the data-idle sensitivity from the Google error budget,
\[
    w_{\rm idle}\simeq 10,
\]
the corresponding sensitivity-weighted contribution is
\[
    C_{\rm mem}
    \simeq
    w_{\rm idle}p_{\rm mem}(\tau_{\rm cyc})
    \simeq
    10\times 1.02\times 10^{-2}
    \simeq
    0.102 .
\]
This is close to the calibrated data-idle contribution reported in the
Google error budget,
\[
    C_{\rm idle}^{\rm data}=0.090 .
\]
Equivalently, relative to the non-idle data-like contribution
\[
    C_{\rm gate}^{\rm data}
    =
    C_{\rm CZ}
    +C_{\rm CZ,xtalk}
    +C_{\rm CZ,leak}
    +C_{\rm SQ}
    =
    0.293,
\]
the coherence-time estimate would give
\[
    \lambda_{\rm mem}
    =
    \frac{C_{\rm mem}}{C_{\rm gate}^{\rm data}}
    \simeq
    \frac{0.102}{0.293}
    \simeq
    0.35 .
\]
By contrast, the direct calibration from the reported data-idle budget
gives
\[
    \lambda_{\rm Google}
    =
    \frac{C_{\rm idle}^{\rm data}}{C_{\rm gate}^{\rm data}}
    =
    \frac{0.090}{0.293}
    \simeq
    0.31 .
\]
Although we derive $\lambda_{\rm Google}$ and $\lambda_{\rm mem}$ using two different approaches, they are consistent with each other.

\end{document}